\documentclass[journal]{IEEEtran}

\usepackage{ifpdf}
\usepackage{cite}
\ifCLASSINFOpdf
   \usepackage[pdftex]{graphicx}
\else
\fi
\usepackage{amsmath}
\usepackage{algorithmic}
\usepackage{fixltx2e}
\usepackage{url}

\usepackage{graphicx}
\usepackage[space]{grffile}

\usepackage{comment}
\usepackage{romannum}
\usepackage{multirow}
\usepackage{subfig}
\usepackage[font=small,labelfont=bf]{caption}
\usepackage{balance}

\newcommand{\subhead}[1]{
	\paragraph{#1}
	}

\begin{document}
\title{Attacking Image Splicing Detection and Localization Algorithms Using Synthetic Traces}

\author{Shengbang~Fang,
        Matthew~C~Stamm,~\IEEEmembership{Member,~IEEE}}%

\maketitle

\IEEEpeerreviewmaketitle

\begin{abstract}
Recent advances in deep learning have enabled forensics researchers to develop a new class of image splicing detection and localization algorithms. These  algorithms identify spliced content by detecting localized inconsistencies in forensic traces using Siamese neural networks, either explicitly during analysis or implicitly during training.  
At the same time, deep learning has enabled new forms of anti-forensic attacks, such as adversarial examples and generative adversarial network (GAN) based attacks.  
Thus far, however, no anti-forensic attack has been demonstrated against image splicing detection and localization algorithms. 
In this paper, we propose a new GAN-based anti-forensic attack that is able to fool state-of-the-art splicing detection and localization algorithms such as EXIF-Net, Noiseprint, and Forensic Similarity Graphs.  This attack operates by adversarially training an anti-forensic generator against a set of Siamese neural networks so that it is able to create synthetic forensic traces.  Under analysis, these synthetic traces appear authentic and are self-consistent throughout an image.   Through a series of experiments, we demonstrate that our attack is capable of fooling forensic splicing detection and localization algorithms without introducing visually detectable artifacts into an attacked image.  Additionally, we demonstrate that our attack outperforms existing alternative attack approaches.  %

\unskip
\end{abstract}


\section{Introduction}
\label{sec:intro}

Digital editing software presents a significant challenge when determining the authenticity of digital images.  In response, media forensics research has been established to combat this problem.  Forensic algorithms operate by exploiting traces left by image editing or an image's source camera.  
In the past, this was largely done by building hand designed models of forensic traces through theoretical analysis~\cite{stamm2013overview, piva2013overview, verdoliva2020overview}.  

Due to advances in deep learning, however, convolutional neural networks (CNNs) have enabled researchers to learn forensic traces directly from data.  This has enabled researchers to build powerful CNN-based  algorithms capable of performing a wide variety of forensic tasks including manipulation detection~\cite{bayar2018constrained, chen2015median, cozzolino2017recasting, BARNI2017153}, camera model identification~\cite{bondi2016first, bayar2017design, tuama2016camera}, and tracing an image's distribution through social networks~\cite{Amerini_2017_WIFS}. 

One important problem in image forensics is detecting and localizing spliced image content.  Image splicing occurs when content from one image is pasted into another.  As a result, the spliced content will have different forensic traces than the rest of the image.  
Since it is infeasible to learn all possible source or editing traces, forensic algorithms operate by searching for localized inconsistencies in the traces within an image.  Currently, state-of-the-art splicing detection and localization techniques use Siamese neural networks, either explicitly in algorithm or as part of learning features.  These include powerful algorithms such as Noiseprint~\cite{cozzolino2019noiseprint}, Forensic Similarity Graphs~\cite{mayer2020exposing}, and EXIF-Net~\cite{huh2018fighting}.

While forensics helps protect against falsified and manipulated images, an intelligent forger will attempt to use anti-forensic countermeasures to fool forensic algorithms.   Several anti-forensic attacks have been proposed in to fool algorithms that %
detect JPEG Compression~\cite{stamm2010anti, fan2014jpeg}, resampling~\cite{Kirchner_2008_TIFS}, median filtering~\cite{wu2013anti,fontani2012hiding}, and evidence of LCA inconsistencies~\cite{Owen2015antilca}.
These attacks typically operate by creating fake forensic traces that match a target forensic algorithm's human-designed model of traces in an unaltered image.   However this design approach is not feasible when attacking deep learning algorithms, which utilize learned models of traces that are not easily interpretable to humans.

The widespread adoption of deep learning in forensics has enabled new forms of anti-forensic attacks such as adversarial examples.  
Adversarial examples operate by exploiting vulnerabilities in deep neural networks~\cite{Akhtar2018survey}.  These attacks create small image perturbations that cause a classifier to misclassify an input. Several algorithms have been proposed to create adversarial examples including the Fast Gradient Sign Method (FGSM)~\cite{goodfellow2015explain}, Project Gradient Descent (PGD)~\cite{madry2018towards}, and Carlini-Wagner (CW) algorithms~\cite{carlini2017towards}.  Additionally, attacks on forensic CNNs have been proposed using adversarial examples~\cite{Barni_2019_ICASSP, Carlini2020CVPRW, rozsa2020adversarial, Guera2017CVPRW, Zhong_2021_TIFS}.

Though adversarial examples are a significant threat to forensic classifiers, they are unlikely to fool forensic splicing detection and localization algorithms in their current form.  Splicing detectors and localizers are not simple classifiers.  They operate by splitting an image into several small analysis blocks, then looking for anomalous forensic traces.  
By contrast, adversarial examples are designed to fool neural networks analyzing one sample at a time.  They will have a difficult time creating consistent forensic traces throughout an entire image since each analysis block must be attacked individually.
Furthermore, they may encounter issues surround misalignment with how the attacker and forensic algorithm divides an image into analysis blocks. 

At the same time, GAN-based attacks have emerged as a new deep learning based anti-forensic threat~\cite{Chen_2018_ICIP, Chen_2019_TIFS, Xie_2021_TCSVT, Kim_2018_SPL, Ding_2021_TM}.  GAN-based attacks do not exploit classifier vulnerabilities. Instead, they use a generator to synthesize fake forensic traces that a forensic neural network will associate with a target class (e.g. the class of unltered images).  The generator is itself a deep neural network that learns to synthesize these traces through adversarial training.  However, GAN-based attacks have so far only been demonstrated against forensic classifiers, not splicing detectors or localizers.

In this paper, we propose a new GAN-based attack capable of fooling forensic splicing detection and localization algorithms based on Siamese neural networks.  
Instead of  independently attacking individual image patches, this attack synthesizes new homogeneous forensic fingerprints directly in a full-sized  image. It does this by passing a spliced image through a convolutional generator that is adversarially trained to synthesize realisitic and self-consistent forensic traces throughtout an image.  A novel, two phase training strategy is proposed to train this attack.
In Phase~1, we pre-train the generator to learn forensic feature embeddings by adversarially training against a forensic CNN. 
In Phase~2, we refine the generator's learned model to create self-consistent forensic traces capable of fooling forensic Siamese networks.

Through a series of experiments, we demonstrate that our new attack can fool state-of-the-art splicing detection and localiztion algorithms including Noiseprint, Forensic Similarity Graphs, and EXIF-Net. These experiments show that our attack can maintain high visual quality while fooling these algorithms.  Furthermore, our experiments demonstrate that our attack outperforms existing anti-forensic and adversarial example based attacks.  Finally, we demonstrate a new misinformation threat that is made feasible by our proposed attack.

We summarize the main contributions of this paper as follows: 
(1) We propose a new attack capable of fooling forensic detection and localization algorithms.  Our attack uses a novel training strategy to enable an adversarially trained generator to synthesize self-consistent fake forensic traces throughout an entire attacked image.  Our training strategy enables our attack to fool multiple forensic algorithms at once using an ensemble Siamese loss term.  Furthermore our attack is generic and reusable, i.e. it can attack any image of any size and does not need to be retrained for each image, unlike adversarial example attacks.  
(2) We provide extensive experimental evaluation of our attack.  We show that our proposed multi-stage training strategy outperforms other alternatives, including directly training against a CNN as is done int the MISLgan attack.  We show that our attack is able to fool forensic Siamese networks operating on a patch level, as well as state-of-the-art forensic detection and localization algorithms such as EXIF-Net, Noiseprint, and Forensic Similarity Graphs.  Furthermore, we show that our attack can transfer to fool algorithms that it was not directly trained against.  Additionally, we show that our attack maintains high image quality in all experiments.  
(3)  We demonstrate that our attack outperforms existing attacks.  Our experiments show that adversarial example attacks are poorly suited to attacking splicing detectors and localizers.  Additionally, we show that our attack outperforms other alternatives that directly attack features such as the LOTS attack.  
(4) Finally, we demonstrate a new misinformation threat enabled by our attack in which real content is made to convincingly appear fake under forensic analysis.  This can potentially be used by an attacker to cast doubt on authentic images.

\section{Image Splicing Detection \& Localization}
\label{sec:related}
Image splicing occurs when content is cut from one image and pasted into another. Prior research has shown that splicing detection and localization algorithms also perform well when localizing image editing or tampering. Many image splicing detection and localization algorithms have been developed using image forensic traces. Recently, several deep learning-based approaches with Siamese structures have achieved much greater success in detection and localizing fake content than other competing approaches. These include algorithms such as EXIF-Net, Forensic Similarity Graphs, and Noiseprint. While each of these algorithms operates differently, all utilize forensic siamese neural networks either explicitly as part of their algorithm or during the training of their feature extractor. Below we provide a brief overview of each of these algorithms.

\paragraph{EXIF-Net} EXIF-Net proposed by Huh~\cite{huh2018fighting} performs image splicing detection and localization. It uses a Siamese network, known as EXIF-Self Consistency Network (EXIF-Net), to determine whether two image patches in an image came from the same source, as defined by an image's metadata. The EXIF-Net is trained to output a score of 1 if two patches came from an image with the same source and 0 from different sources. The splicing detection and localization are made by using EXIF-Net to compare all pairs of sampled patches over the full-size image. Average self-consistency scores are used to make detection decisions. For localization, the sampled patches are clustered using mean shift. Then a response map is built to highlight the spliced content.

\paragraph{Forensic Similarity Graph} Forensic Similarity Graph proposed by Mayer and Stamm~\cite{mayer2020exposing} is designed to detect and localize image tampering, including splicing, editing and etc. In the algorithm, an image is first sampled in a grid. The forensic traces in these sampled patches are compared to one another using a "forensic similarity metric"~\cite{Mayer2020TIFS}, which is measured by a Siamese network. This Siamese network is trained to determine if two image patches came from the same source camera. Research by these authors has also shown that training the Siamese network in this manner can also apply to identify differences in editing and other processing. After that, a forensic similarity graph is built by treating each patch as a node and the forensic similarity score between two patches as their edge weight. Finally, tampering is detected and localized by measuring the community structure of the graph. 

\paragraph{Noiseprint} Noiseprint proposed by Cozzolino~\cite{cozzolino2019noiseprint} is designed to perform image forgery localization. It should be noted that Noiseprint does not explicitly perform forgery detection. Instead, it assumes that an investigator is presented with a falsified image, then Noiseprint is used to localize the falsified content. Noiseprint contains a convolutional feature extractor that is designed to encode an image's source camera model's ``fingerprints''. This feature extractor is trained in a Siamese structure, then only the feature extractor is retained after training. Forgery localization is performed by sampling an image's ``Noiseprint'' into small patches, then a Gaussian Mixture Model is used to cluster the image patches according to the histogram statistics from each patch.

\section{Attack Requirements}
\label{sec:problem}

In this section, we discuss the goals and requirements of our anti-forensic attack. We first introduce the goals of any anti-forensic attack, 
then discuss goals and requirements specific to attacking image splicing detectors and localizers. 
Additionally, we briefly discuss the knowledge available to the attacker.

For any anti-forensic attack, a successful attack should fulfill at the following requirements: 
(1) The attack should fool the forensic system. This can have different meaning in different contexts, and can be accomplished in multiple different ways.  For example, a classifier can be fooled by  mimicking the forensic trace associated with an authentic image or by introducing adversarial perturbations to induce errors in classifier.  
(2) The attack should not be perceptible to a human.  This means that the attack should introduce no visible noise or distortions that would make an image appear implausible.  We note that since the attacker will only release the attacked image, they do not have to make the attacked image match the unattacked version, just that it is plausibly authentic.  
(3) The attack should be generic, i.e.  it should be able to attack any image of any size.

When attacking an image splicing detector or localizer, it is necessary to clarify the meaning of a ``successful'' attack.  When fooling a detector, an attack should make the detector unable to tell the difference between real and falsified images.  This means the detectors performance should be reduced to a randome decision or worse, e.g. the area under the ROC curve (AUC) is reduced to 0.5 or less.  When fooling a localizer, an attack should make the localizer unable to correctly predict the falsified region of an image. This corresponds to causing substatial reductions in localization scores such as the Matthew's Correlation Coefficient (MCC) or F1 scores that measure localiztion correspondence with a ground truth mask.

We also note that ideally, the attack should simultaneously fool the detector and localizer. The ultimate goal of the attack is to make a falsified image look real. If the attack fools the localizer but does not fool the detector, the attack still fails.  This is because detector can still reliably identify the image is a forgery. 
If the attack fools the detector but does not fool the localizer, the attack is still potentially unsuccessful.  If the localizer predicts a very meaningful forgery location (i.e. it corresponds to an object or person in an image)  this might alarm the investigator.

Finally, we note that the knowledge available to the attacker also 
influences the attack's success.
In this paper, we consider two levels of information available to the attacker: white box and zero knowledge attacks .
In the white box scenario, the attacker has full information about the  forensic algorithm under attack.  This means that the attacker can train their attack directly against the target forensic algorithm.  
In the zero knowledge scenario, the attacker does not have any information about the forensic algorithm  they wish to attack.  We note that this is different from a black box attack in which an attacker can query the forensic algorithm and observe its output.  In the zero knowledge scenario, an attack must rely entirely on attack transferability.  This means the attack is trained directly against one set of forensic algorithms, but must also fool other algorithms without any additional training.

In this paper, we focus primarily on white box attacks.  In this scenario, we launch our attacks directly against the EXIF-Net and Forensic Similarity Graph detection and localization algorithms. %
Additionally, we demonstrate that a zero knowledge attack is feasible by showing that our attack can also transfer to fool the Noiseprint forgery localization algorithm without explicitly training against it.

\section{Proposed Attack}
\label{sec:proposed}
\begin{figure}[t]
\centering
\includegraphics[scale=0.27]{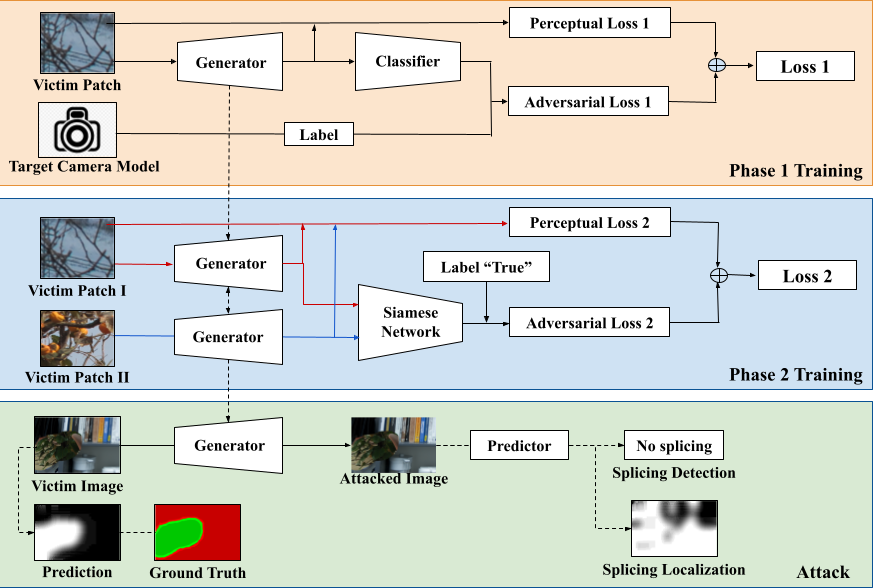}
\caption{Overview of our proposed attack. 
In Training Phase 1, the attack generator $G$ is adversarially trained against a camera model classifier $C$. 
In Training Phase 2, the pre-trained generator $G$ is adversarially trained to synthesize traces to fool forensic Siamese networks.
The attack is deployed by passing a spliced image through the fully trained attack generator $G$.}
\label{fig:Pipeline}
\end{figure}

In this section, we describe the details of our proposed attack based on a generative adversarial network. 
This new attack uses a trained generator $G$ to synthesize fake forensic traces in the image.
These synthetic forensic traces are designed to mimic real traces associated with an unaltered image and be uniformly consistent throughout the attacked image. A special training framework consisting of two phases is used to train our anti-forensic generator.
In Phase 1, we pre-train the generator to attack a camera model classifier. This is done to initialize our attack and let the generator have a good starting point when learning to synthesize successful attack features in Phase 2. Specifically, we attack camera model features in Phase 1 because prior research has shown that camera model features are transferable to other forensic tasks~\cite{mayer2018transfer}. 
Then in Phase 2, we train the initialized generator $G$ to attack a forensic siamese network. 

When launching our attack, we use the generator $G$ after Phase 1 \& 2 training is completed to modify the image. This is done by simply passing a full-size image through the trained generator. The generator is only trained once, then is reused to attack any spliced image. 
An overview of our attack is depicted in Fig.\ref{fig:Pipeline}.

\subsection{Generator Architecture}
The architecture of our generator $G$ is shown in the Table.~\ref{table: Models}. In order to make the generator capable of synthesizing forensic traces to avoid visible up-sampling artifacts~\cite{Zhang2019wifs}, we use a fully convolutional architecture, where each convolutional layer uses a stride of 1. In detail, the architecture first uses one $3\times3$ convolutional layer, %
which is followed by five residual blocks~\cite{he2016deep}.
Each residual block consists of two convolutional layers with 64 $3\times3$ kernels with stride one and a skip connection. 
The use of residual blocks is to protect against the vanishing gradient problem.
After that, we use a convolutional layer with three kernels with size $3\times3$ and stride 1 to achieve three-channel (i.e. RGB) output images. We use a ReLU activation function after all convolutional layers in the generator architecture.

\begin{table}
\caption{The structure of attack generator}
\label{table: Models}
\small
\centering
\setlength{\tabcolsep}{3 mm}
\renewcommand{\arraystretch}{1.2}
\begin{tabular}{|c|c|c|c|}
\hline
 \multicolumn{4}{|c|}{Generator Configuration}\\
 \hline
 Name & Layer & Kernel Size & Channels\\
 \hline 
 Input Layer & 1 & 3$\times$3 & 64\\
 \hline 
 \multirow{2}{*}{ResNet Block-\Romannum{1}} & 2 & \multirow{2}{*}{3$\times$3} & \multirow{2}{*}{64} \\
 \cline{2-2} & 3 & & \\
 \hline
 \multirow{2}{*}{ResNet Block-\Romannum{2}} & 4 & \multirow{2}{*}{3$\times$3} & \multirow{2}{*}{64} \\
 \cline{2-2} & 5 & & \\
 \hline
 \multirow{2}{*}{ResNet Block-\Romannum{3}} & 6 & \multirow{2}{*}{3$\times$3} & \multirow{2}{*}{64} \\
 \cline{2-2} & 7 & & \\
 \hline
 \multirow{2}{*}{ResNet Block-\Romannum{4}} & 8 & \multirow{2}{*}{3$\times$3} & \multirow{2}{*}{64} \\
 \cline{2-2} & 9 & & \\
 \hline
 \multirow{2}{*}{ResNet Block-\Romannum{5}} & 10 & \multirow{2}{*}{3$\times$3} & \multirow{2}{*}{64} \\
 \cline{2-2} & 11 & & \\
 \hline
 Reduction & 12 & 3$\times$3 & 3 \\
 \hline
\end{tabular}
\end{table}

\subsection{Training Phase 1}
Our generator $G$ is pre-trained to synthesize forensic traces associated with different camera models in Phase 1. We do this because prior research has shown camera model features are transferable to other forensic tasks~\cite{mayer2018transfer}, and because we have experimentally observed an increase in attack performance by utilizing this training phase.

To begin, we first construct a camera model classifier $C(\cdot)$ that we will adversarially train $G$ to fool. In practice, we use the MISLNet CNN for $C(\cdot)$~\cite{bayar2018constrained}. Next, we randomly pick one of the $N$ different camera models that $C$ is trained to identify as the attack's target camera model.  
We then train $G$ to synthesize forensic traces associated with the target camera model by adversarially training $G$ against $C$. 
This training process can force the generator $G$ to learn to synthesize new forensic traces of the target camera model. To achieve a generator with a good attack capability and to maintain good image quality, we set the Phase 1 loss function $L_{G1}$ for the generator $G$ to be a combination of classification loss $L_C$ and Phase 1 perceptual loss $L_{P1}$

\begin{equation}
\label{loss_1}
L_{G1} = \alpha L_{C} + L_{P1}.
\end{equation}
The term $\alpha$ %
is a scalar used to balance the weight between %
the perceptual loss $L_{P1}$ and the classification loss $L_{C}$.

The classification loss measures the confidence of the attacked image being from the target camera model. We define the classification loss to be the multi-class cross-entropy between $C$'s output and the target output
\begin{equation}
\label{Classification loss}
L_{C} = -\sum_{k=1}^{n} t_{k} log(C(G(I))_{k}),
\end{equation}
where \textbf{t} is a one-hot vector that has 1 in the index k of the target camera model. Using this loss function, we can force our generator $G$ to learn to synthesize the forensic traces of camera models.

The perceptual loss $L_{P1}$ is to constrain the distortion on the image introduced by our generator $G$. We use the $L_1$-norm as the loss to measure the difference between the generator's input image $I$ and the output image $G(I)$.  The perceptual loss of our generator $L_{P1}$ is defined as 
\begin{equation}
\label{perceptual loss}
L_{P1} = \frac{1}{w \times h \times c} \sum_{i=1}^{w} \sum_{j=1}^{h} \sum_{k=1}^{c}|I_{i,j,k}-G(I)_{i,j,k}|,
\end{equation}
where i, j are the locations of the pixels in the image, $w$ and $h$ are the image's width and height, and $c$ is the number of color channels in the image.  %

\subsection{Training Phase 2}
In Phase 2, we train the initialized generator $G$ in Phase 1 to attack the targeted forensic Siamese network. The targeted forensic Siamese network can be any Siamese network that accepts a pair of image patches as the input, then produces a confidence score measuring similarity of the forensic traces in these patches as the output.  For example, both the Forensic Similarity Graph algorithm~\cite{mayer2020exposing} and EXIF-Net~\cite{huh2018fighting} use forensic Siamese networks as part of their algorithm. 

Similar to Phase 1 training, we define the loss function $L_{G2}$ for training  $G$ during Phase 2 to be a combination of a similarity loss $L_{S}$ and a perceptual loss $L_{P2}$:
\begin{equation}
\label{loss_2}
L_{G2} = L_{S} + L_{P2},
\end{equation}

In this phase, we define $L_{P2}$ as the mean squared error (MSE) between the input and output image patches
\begin{equation}
\label{perceptual loss}
L_{P2} = \frac{1}{w \times h \times c} \sum_{i=1}^{w} \sum_{j=1}^{h} \sum_{k=1}^{c}||I_{i,j,k}-G(I)_{i,j,k}||^2.  
\end{equation}
We use MSE as the perceptual loss in this phase because we have experimentally found that it achieves better attacked image quality.

The similarity loss is designed to measure how similar the forensic traces are between two attacked image patches (i.e. if the attacked patches are able to fool a forensic Siamese network into believing that they contain the same forensic traces).  Since multiple forensic Siamese networks exist, we define our similarity loss to be a linear combination of losses associated with different Siamese networks.  Specifically, we define the $L_{S}$ as \begin{equation}
\label{similarity_term}
L_{S} = \sum_{n=1}^{N}\beta_{n}L_{S_{n}},
\end{equation}
where the scalars $\beta_n$ control the weight associated with the $n^{th}$ Siamese network's individual similarity loss $L_{S_{n}}$.  If $n$ is greater than one, this effectively corresponds to training $G$ against an ensemble of targeted forensic Siamese networks.

Assume each targeted Siamese network $S_{n}(\cdot, \cdot)$ used during training  outputs a scalar as the measurement of the similarity between the forensic traces in two input image patches $I_{1}$ and $I_{2}$. We define the individual similarity loss of the $n^{th}$ targeted forensic Siamese network used during training as 
\begin{equation}
\label{similarity_loss}
L_{S_{n}} = -log(S_{n}(G(I_{1}), G(I_{2})).
\end{equation}

\subsection{Deploying the Attack}

Once Phase 1 \& 2 training is finished, the generator $G$ is able to synthesize new forensic traces such   that any two attacked image patches should be classified to be "same" by each targeted Siamese network. 

To launch the trained attack, we retain only the generator $G$ and discard other components.  An image $I$ is attacked by passing it through the trained anti-forensic generator $G$.  This will result in an attacked image $G(I)$ that has the same size and visual content as $I$, but with falsified forensic traces.  Since the generator has a fully convolutional architecture, it accepts images of any size as an input.  As a result,  we can directly feed the whole image $I$ to the generator and obtain the output as the attacked image. If the image is too large for the computational resources available to the attacker, we can also divide the full-size image into smaller patches and use the generator $G$ to attack each individually, then stitch together the attacked image patches to achieve the full-size attacked image.

\section{Experimental Results}
\label{sec:experiments}

In this section, we present a series of experiments designed to evaluate the performance of our proposed attack.  We begin by providing specific details of how our attack was trained for these experiments.  After this, we evaluate our attack's performance directly against forensic Siamese networks, as well as our attack's ability to defeat splicing detection and localization algorithms.  Finally, we compare our attack's performance against other attacks.

\subsection{Training the Attack} 

\subsubsection{Attack Training Dataset}
\label{experiments_training}

To conduct our experiments, we utilized an image dataset consisting of unaltered images manually captured by our lab using 23 unique camera models.  We note, this image dataset has been used to train and evaluate several previous publications~\cite{bayar2018constrained, Chen_2018_ICIP, Chen_2019_TIFS, Mayer2020TIFS}.   This larger dataset was subdivided into two distinct groups: Group A which consisted of images captured by 18 different ``seen'' camera models and Group B which consisted of images captured by 5 ``unseen'' camera models.  
``Seen'' camera models are those that are encountered by our attack during its training phase.  No images from  ``unseen'' camera models are ever used while training our attack, therefore it has not opportunity to learn a model of forensic traces from these cameras.

Images from Group A were divided into distinct training, validation and testing sets.  A set of non-overlapping color image patches of size $128\times128$ pixel patches were extracted from each image.  For each camera model, a total of 30,000 patches were retained for the training set, 4,000 patches were retained for the validation set, and 8,000 patches were retained for the testing set.  This resulted in a training set of 540,000 patches, a validation set of 72,000 patches, and a testing set of 144,000 patches for Group A.

Since Group B consists of ``unseen'' camera models, only a testing set of patches was made for each camera model.  A set of 8,000 patches was retained for each of the five camera models in Group B using the same procedure as outlined above (for Group A).  This resulted in a testing set consisting of 40,000 in total from Group B.

\subsubsection{Attack Training Settings}

\paragraph{Training  $C$} First, we trained the classifier $C$ in Phase 1 to distinguish between the 18 camera models in Group A.  The architecture of $C$ was chosen to be a full-color version of the MISLNet classifier~\cite{bayar2018constrained} due to its strong reported camera model identification performance.  $C$ was trained using the $128\times 128$ pixel patches from the group A training set for 10 epochs, with a learning rate of 1e-3 and, decay for every 3 epochs.  We verified the performance of $C$ by using it to identify the source camera model of each image patch in Group A's test set.  The trained classifier achieved an average accuracy of 97.3\% which is consistent with the results in the original publication~\cite{bayar2018constrained}. 

\paragraph{Training $G$ - Phase 1} Next, we trained the generator $G$ in Phase 1 to directly fool the classifier $C$. Again, we still only used the group A training set for training. We randomly picked one camera model as the target model to provide the label $t_{k}$ for the cross-entropy classification loss $L_{C}$. We also set the coefficient $\alpha$ to be 20 for shortening the training procedure and maintaining acceptable image quality. We used the Adam Optimization method, with a starting rate of 1e-4 and decay in half for every three epochs. We use a batch size of 32 to train the $G$ for 10 epochs.

\paragraph{Training $G$ - Phase 2} Finally, we trained $G$ to fool forensic Siamese networks in Phase 2 of its training.  In these experiments, we considered two different forensic Siamese ntworks: $S_{1}$ is the Forensic Similarity Graph (FSG) proposed by Mayer and Stamm~\cite{mayer2020exposing} and $S_{2}$ the EXIF-Self Consistency Network (EXIF-Net) proposed by Huh~\cite{huh2018fighting}.  Both networks use a common input patch size of $128 \times 128$  pixels.  
To compare our attack on the two $S_{n}$s, we trained  $G$ using four different strategies: training  $G$ using only $S_{1}$, training  $G$ using only $S_{2}$, training using both $S_{n}$s but without training Phase 1, and training using both $S_{n}$s with Phase 1 training.  When training using both $S_{n}$,  we set $\beta_{1} = 800$ and  $\beta_{2} = 30$.  These $\beta_{n}$s were chosen to yield strong attack performance after conducting several small-scale preliminary experiments.

\begin{table*}[h]
\caption{Image patch level performance of our proposed attack}
\label{table: PatchPerformance}
\small
\centering
\setlength{\tabcolsep}{1.5 mm}
\renewcommand{\arraystretch}{1.2}
\begin{tabular}{|c|c|c|c|c|c|c|c|c|c|}
\hline
 \multirow{2}{*}{Training $S_{n}$} & \multirow{2}{*}{Evaluation Method} & \multicolumn{4}{c|}{Seen Models} & \multicolumn{4}{c|}{Unseen Models}\\
 \cline{3-10}
 & & FMR & SAR & m-PSNR & m-SSIM & FMR & SAR & m-PSNR & m-SSIM\\
 \hline
 FSG & \multirow{4}{*}{FSG} & 6.9\% & 99.9\% & 45.2 & 0.991 & 7.2\% & 99.9\% & 45.2 & 0.991 \\
 \cline{1-1}\cline{3-10}
 EXIF-Net & & 6.9\% & 34.9\% & 42.5 & 0.981 & 7.2\% & 35.8\% & 42.5 & 0.981 \\
 \cline{1-1}\cline{3-10}
 Both, No Init & & 6.9\% & 99.9\% & 40.9 & 0.975 & 7.2\% & 99.9\% & 40.9 & 0.975 \\
 \cline{1-1}\cline{3-10}
 Both & & 6.9\% & \bf{99.9\%} & 41.3 & 0.976 & 7.2\% & \bf{99.9\%} & 41.4 & 0.976\\
 \hline 
 FSG & \multirow{4}{*}{EXIF-Net} & 13.6\% & 15.6\% & 45.2 & 0.991 & 14.3\% & 16.4\% & 45.2 & 0.991 \\ 
 \cline{1-1}\cline{3-10}
 EXIF-Net & & 13.6\% & 96.3\% & 42.5 & 0.981 & 14.3\% & 96.2\% & 42.5 & 0.981 \\
 \cline{1-1}\cline{3-10}
 Both, No Init & & 13.6\% & 97.0\% & 40.9 & 0.975 & 14.3\% & 97.1\% & 40.9 & 0.975 \\
 \cline{1-1}\cline{3-10}
 Both & & 13.6\% & \bf{97.4\%} & 41.3 & 0.976 & 14.3\% & \bf{97.5\%} & 41.4 & 0.976\\
 \hline
\end{tabular}
\end{table*}

When training $G$ to fool each Siamese network, we continue to use patches from the training set of Group A.  To generate each mini-batch of image patch pairs to use during training, we first randomly choose 48 image patches.  Next, we randomly draw 32 pairs of image patches from this set (with replacement) to produce a set of input patches for each $S_{n}$ during attack training. We use 1e-4 as the starting learning rate and decay in half for every 3 epochs. We train the generator $G$ for a maximum of 20 epochs and have an early stop when the successful attack rate~\eqref{SAR} on the validation set begins decreasing. We note that we do not have any specific target camera models in this phase because the $G$ is to make every pair of image patches to be classified as the same by the $S_{n}$s.

\subsection{Directly Attacking on Siamese Networks}
\label{sec:result_dr}

In our first experiment, we evaluated our attack's ability to directly fool a forensic Siamese network, such as Forensic Similarity Graph (FSG) or EXIF-Net.
 These Siamese networks can be used to determine if a pair of image patches come from the same source. They will output a score $>0.5$ for pairs from the same source and  a score $<0.5$ for pairs from different sources.  Here, we evaluated the ability of our proposed attack to fool these Siamese networks into 
determining that two patches from a different camera model came from the same source (i.e. produce an output score $>0.5$ )

\subhead{Dataset}

For this experiment, we evaluated our attack's ability to fool a forensic Siamese network given patches from both the `seen' and `unseen' test sets described above in Section~\ref{experiments_training}.  
To generate pairs of patches from different sources, we first shuffled the test set of each camera model.  We then randomly selected 8,000 pairs of image patches for every combination of two camera models.   This resulted in 1,224,000 pairs of patches from `Seen' models and 80,000 pairs from `Unseen' models.

\subhead{Patch-Wise Attack Evaluation Metrics}

We measure our attack's ability to fool a forensic Siamese network using the successful attack rate (SAR).  In this experiment, SAR measures the likelihood that two attacked patches from different sources will be misclassified by $S_n$ as originating from the same source.  Specifically,
\begin{equation}
\label{SAR}
\footnotesize
SAR = P[S_{n}(G(X_1), G(X_2)) \leq 0.5 | X_1 \in \mathcal{I}_j,\; X_2 \in \mathcal{I}_k,\; j\neq k]
\end{equation}
where $\mathcal{I}_i$ and $\mathcal{I}_j$ are the set of images from cameras $i$ and $j$.

It is important to place SAR in the context of a Siamese network's performance before an attack is launched.  We do this by comparing the SAR to the false match rate (FMR), which we define as the likelihood that an two unattacked patches from different sources are misclassified as originating from the same source by $S_n$.  Specifically, 
\begin{equation}
\label{Rorg}
\footnotesize
FMR = P[S_{n}( X_1, X_2 ) \leq 0.5|  X_1 \in \mathcal{I}_j,\; X_2 \in \mathcal{I}_k,\; j\neq k].
\end{equation}

\subhead{Image Quality}
In addition to fooling the Siamese network, our attack should not leave behind traces that are perceptibly visible to humans.  As a result, is important to measure the distortion introduced into an attacked image.  We measure the distortion introduced into the attacked image using two metrics: the mean PSNR (m-PSNR) and the mean SSIM (m-SSIM) between original and attacked image patches over the evaluation database. 
We note that in reality, an investigator would never have access to the original image because the attacker would only distribute the attacked image.  As a result, the ultimate goal of the image quality is to make synthesized forensic traces invisible, not to achieve very high m-PSNR and m-SSIM scores.  Instead, these scores act as a quantifiable measure of how noticeable distortions introduced by our attack are.  We heuristically set the  target m-PSNR to be greater or equal to 40 and the target m-SSIM to be greater or equal to 0.96.  It is difficult or impossible for a human to distinguish between the original and distorted version of an image above these levels.  This is a more stringent criteria than requiring that the attack is not noticed, but is easier to measure.

\subhead{Results}

We attacked each of the 1,224,000 pairs of patches from `Seen' camera models in the testing set of Group A as well as the 80,000 pairs of patches from the `Unseen' camera models in the testing set of Group B using our proposed attack.  We doing this, we compared outcomes achieved using four different training strategies (train purely on FSG, on EXIF-Net, train FSG and EXIF-Net both without Phase 1, and train FSG and EXIF-Net both with Phase 1) of our attack. 
The results of this experiment are shown in Table~\ref{table: PatchPerformance}, which separately presents attack results for both `Seen' and `Unseen' camera models.  The best SAR scores of each evaluation method are in boldface.

From Table~\ref{table: PatchPerformance}, we can see that our attack is most successful when training using the `Both' strategy.  In this case, our attack achieves a $SAR=99.9\%$ for patches from both `Seen' and `Unseen' camera models when attacking  FSG.  Similarly, our attack achieves an $SAR=97.4\%$ for `Seen' camera models and $SAR=97.5\%$ for `Unseen' camera models when attacking EXIF-Net.  This result demonstrates that our attack can successfully fool these forensic Siamese networks.  We note that high SARs are also achieved when the attack is trained  and launched against the same Siamese network. However, the attack trained only against FSG has difficulty transferring to attack EXIF-Net and vice-versa.  A potential reason could be that  FSG only focuses on capturing and comparing camera model forensic traces, while  EXIF-Net attempts to capture  camera model traces and other EXIF metadata-related information.  We note that training our attack against both Siamese networks simultaneously overcomes this problem.
Additionally, when comparing results for `Both' (i.e. training with Phase 1 initialization) and `Both, No Init' (i.e. training without initialization), 
we can see that utilizing Training Phase~1 yields both higher SAR and higher image quality for our attack.

Finally, we note that our attack maintains a high image quality.  For the most successful attack trained using the `Both' training strategy, we are able to achieve a mean PSNR at or above 41.3 and a mean PSNR at or above 0.975 for both `Seen' and `Unseen' models when attacking both Siamese networks.  This suggests that our attack will not leave behind traces that are visible to the human eye.

\subsection{Attacking  Image Splicing Detection}
\label{sec:result_det}

Next, we evaluated our attack performance when attacking splicing detection algorithms based on Siamese networks.  For this experiment, we attacked the  FSG %
and EXIF-Net algorithms.
Both algorithms operate by first dividing an image into potentially overlapping analysis blocks, then use a Siamese network to compare forensic traces between pairs of blocks.  Next, the Siamese network outputs are used to build a pair-wise affinity matrix, which is provided as input to a clustering algorithm to perform detection.  We note here that we do not evaluate performance against the Noiseprint algorithm in this experiment because Noiseprint is only designed to perform localization~\cite{cozzolino2019noiseprint}.

\subhead{Dataset}
To evaluate the performance of the proposed attack, we used the Columbia~\cite{hsu06ICME}, Carvalho (DSO-1)~\cite{Carvalho_2013_TIFS} and (Korus) Realistic Tamper~\cite{Korus2016TIFS} databases. The Columbia database contains 183 authentic images and 180 simple spliced images. The DSO-1 database consists of 100 authentic and 100 visually realistic spliced images. The Korus database contains 220 authentic and 220 visually realistic falsified images. On average, the Columbia database has the largest splicing region but with the smallest image sizes. The DSO-1 and Korus databases have larger images and smaller splicing regions.

\subhead{Splicing Detection Evaluation Metrics}
We quantified the performance of %
the FSG and EXIF-Net 
detection algorithms both before attack (i.e. Baseline performance) and after attack using  mean average precision (mAP).  The mAP is equivalent to the area under the ROC curve (AUC) and is a standard metric for assessing splicing detection performance. 
We note that the mAP has three critical levels for  splicing detection: (1) When the mAP = 1, the detector can always successfully detect the falsified image; (2) When  mAP = 0.5, the detector can not discriminate between the spliced and the authentic images and can only make a random guess; and (3) When  mAP = 0, the current detector always predicts the forged images to be authentic and yields convincingly wrong decisions. Both (2) and (3) yield successful attacks since, in both scenarios, the detector is not reliable.  Additionally, we also report the m-PSNR and m-SSIM, as mentioned before, over the attacked splicing images in the evaluation dataset.

\begin{table}[t]
\caption{Splicing detection performance, measured by m-AP}
\label{table: detection}
\small
\centering
\setlength{\tabcolsep}{1.5 mm}
\renewcommand{\arraystretch}{1.2}
\begin{tabular}{|c|c|c|c|c|}
\hline
 \multirow{2}{*}{Training $S_{n}$} & \multirow{2}{*}{\shortstack{Detection\\Method}} & Columbia & DSO-1 & Korus\\
 \cline{3-5}
  & & \multicolumn{3}{c|}{m-AP}\\
 \hline
 Baseline & \multirow{6}{*}{FSG} & 0.94 & 0.93 & 0.67 \\
 \cline{1-1}\cline{3-5} 
 CNN $C(\cdot)$ Only & & 0.82 & 0.76 & 0.62 \\
 \cline{1-1}\cline{3-5} 
 FSG & & 0.46 & 0.14 & 0.10 \\
 \cline{1-1}\cline{3-5} 
 EXIF-Net & & 0.89 & 0.89 & 0.65 \\
 \cline{1-1}\cline{3-5} 
 Both, No Init & & 0.47 & \bf{0.05} & \bf{0.06} \\
 \cline{1-1}\cline{3-5} 
 Both & & \bf{0.46} & 0.08 & 0.15 \\
 \hline
 Baseline & \multirow{6}{*}{EXIF-Net} & 0.98 & 0.76 & 0.54 \\
 \cline{1-1}\cline{3-5} 
 CNN $C(\cdot)$ Only & & 0.98 & 0.84 & 0.52 \\
 \cline{1-1}\cline{3-5} 
 FSG & & 0.97 & 0.85 & 0.51 \\
 \cline{1-1}\cline{3-5} 
 EXIF-Net & & 0.57 & \bf{0.07} & 0.11\\
 \cline{1-1}\cline{3-5} 
 Both, No Init & & \bf{0.32} & 0.13 & 0.09 \\
 \cline{1-1}\cline{3-5} 
 Both & & 0.38 & 0.08 & \bf{0.09} \\
 \hline
\end{tabular}
\end{table}

\begin{table}[t]
\caption{Image quality over three evaluation databases}
\label{table: quality}
\small
\centering
\setlength{\tabcolsep}{1.0 mm}
\renewcommand{\arraystretch}{1.2}
\begin{tabular}{|c|c|c|c|c|c|c|}
\hline
 \multirow{2}{*}{Training $S_{n}$} & \multicolumn{2}{c|}{Columbia} & \multicolumn{2}{c|}{DSO-1} & \multicolumn{2}{c|}{Korus}\\
 \cline{2-7} &PSNR&SSIM&PSNR&SSIM&PSNR&SSIM\\
 \hline
 CNN $C(\cdot)$ Only & 46.87 & 0.994 & 45.59 & 0.990 & 45.78 & 0.992 \\ 
 \hline
 FSG & 44.75 & 0.987 & 43.74 & 0.984 & 43.98 & 0.985 \\
 \hline
 EXIF-Net & 42.84 & 0.981 & 41.79 & 0.976 & 41.61 & 0.980 \\
 \hline
 Both, No Init & 41.63 & 0.975 & 40.86 & 0.967 & 40.97 & 0.973 \\
 \hline
 Both & 42.08 & 0.976 & 40.98 & 0.968 & 40.84 & 0.973 \\
 \hline
\end{tabular}
\end{table}

\paragraph{Results} 
To evaluate our attack's performance against splicing detection algorithms, we launched our attack against the forged images in all three of the databases described above.  Next, we performed splicing detection on using both the FSG and EXIF-Net algorithm and recorded the attacks performance against each algorithm.  This experiment was repeated using the four different training strategies examined previously.

The results of this experiment are listed in Table~\ref{table: detection}, which displays our attacks performance using the proposed training strategy as well as four other alternatives.  Here, we note that the training strategy ``CNN $C(\cdot)$ Only'' uses only Phase 1 training in a manner similar to the MISLgan attack proposed previously in literature~\cite{Chen_2018_ICIP, Chen_2019_TIFS}.  
The lowest m-AP scores (i.e. the best attack performance) of for each dataset and detection algorithm paring are shown in bold.

From Table.~\ref{table: detection}, we can see that when attacking each algorithm using either `Both' training strategy (both with and without Stage 1 initialization), we are able to achieve low m-APs scores.  For the DSO-1 and Korus databases, we were typically able to drop m-AP scores to below 0.10, i.e. we were able to produce very convincingly wrong decisions.  For the Columbia dataset, we were able to drop m-AP scores below 0.5, i.e. reduce the performance of the detector to a random decision or worse.  
We note that when launching our attack against these detectors, Phase-1 initialization seems less important.  It does, however, appear to be important to train our attack against multiple forensic Siamese networks.  Specifically, training using either of our `Both' strategies typically outperforms an attack trained against a single forensic Siamese network, even when the attack is trained explicitly against the $S_n$ that will be used to perform detection.  We note that in subsequent sections, Phase 1 training will be experimentally shown to be important for fooling localization algorithms.
We also note that directly training the generator against a CNN  (i.e. ``CNN $C(\cdot)$ Only''), as is done in prior GAN-based attacks designed to fool forensic classifiers, is largely unsuccessful.  These results demonstrate the need for the multi-phase training protocol proposed in this work.

Additionally, we also show the attacked image quality for the Columbia, Carvalho DSO-1, and Korus databases in Table.~\ref{table: quality}. The attacked images from all three databases have mean-PSNRs higher than 40 and the mean-SSIMs higher than 0.96. These results show that the distortion introduced by our proposed attack are invisible to human eyes. Therefore, an investigator cannot tell if an image is attacked through visual inspection.

\subsection{Attacking  Image Splicing Localization}
\label{sec:result_loc}

\begin{figure*}[!h]
\centering
\setlength{\fboxsep}{0pt}
\begin{subfloat}[Spliced Image\label{spliceimage}]
    {\fbox{\includegraphics[width=0.19\textwidth]{{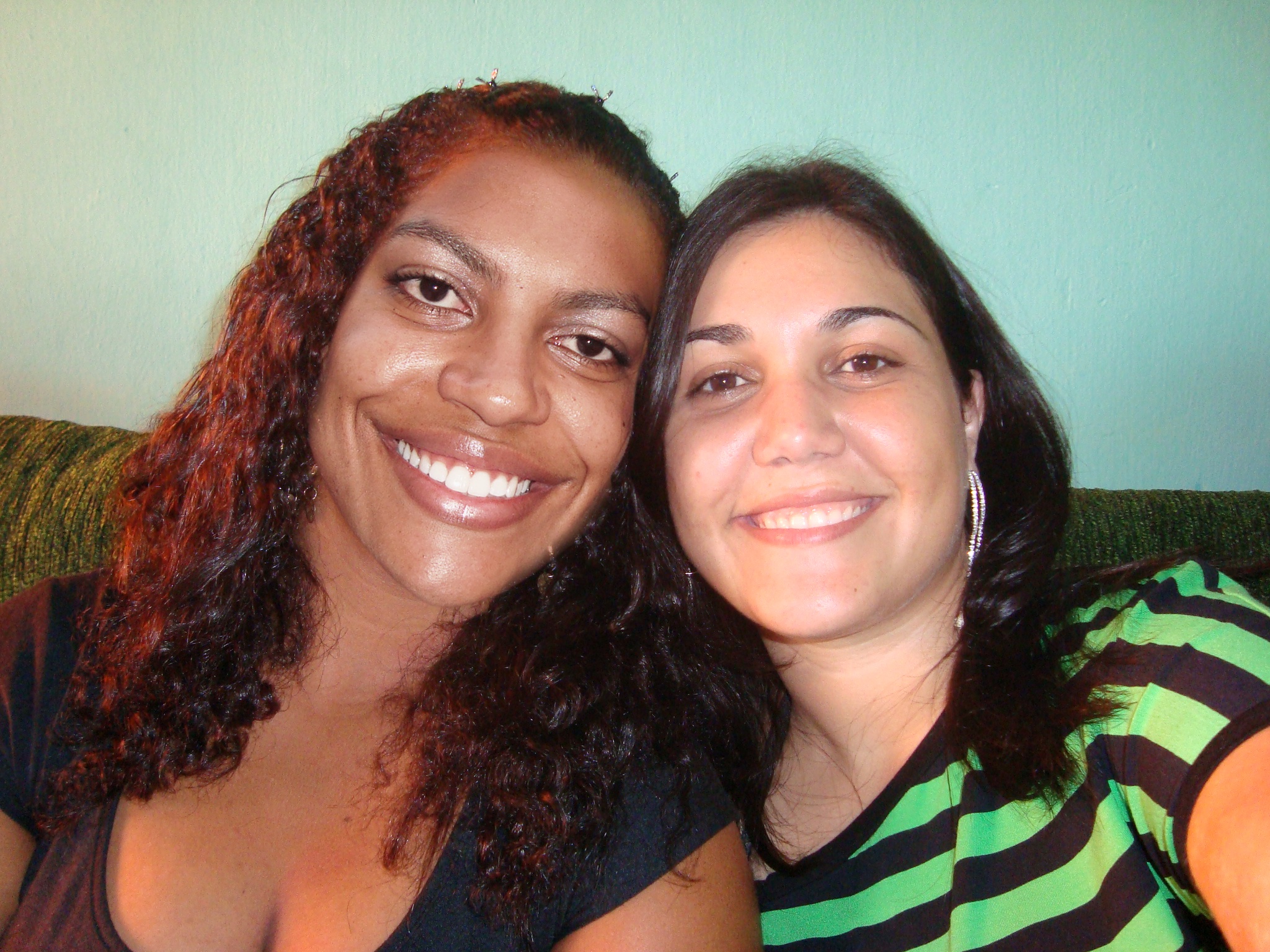}}}}
\end{subfloat}
\begin{subfloat}[Ground Truth Mask\label{grondtruth}]
    {\fbox{\includegraphics[width=0.19\textwidth]{{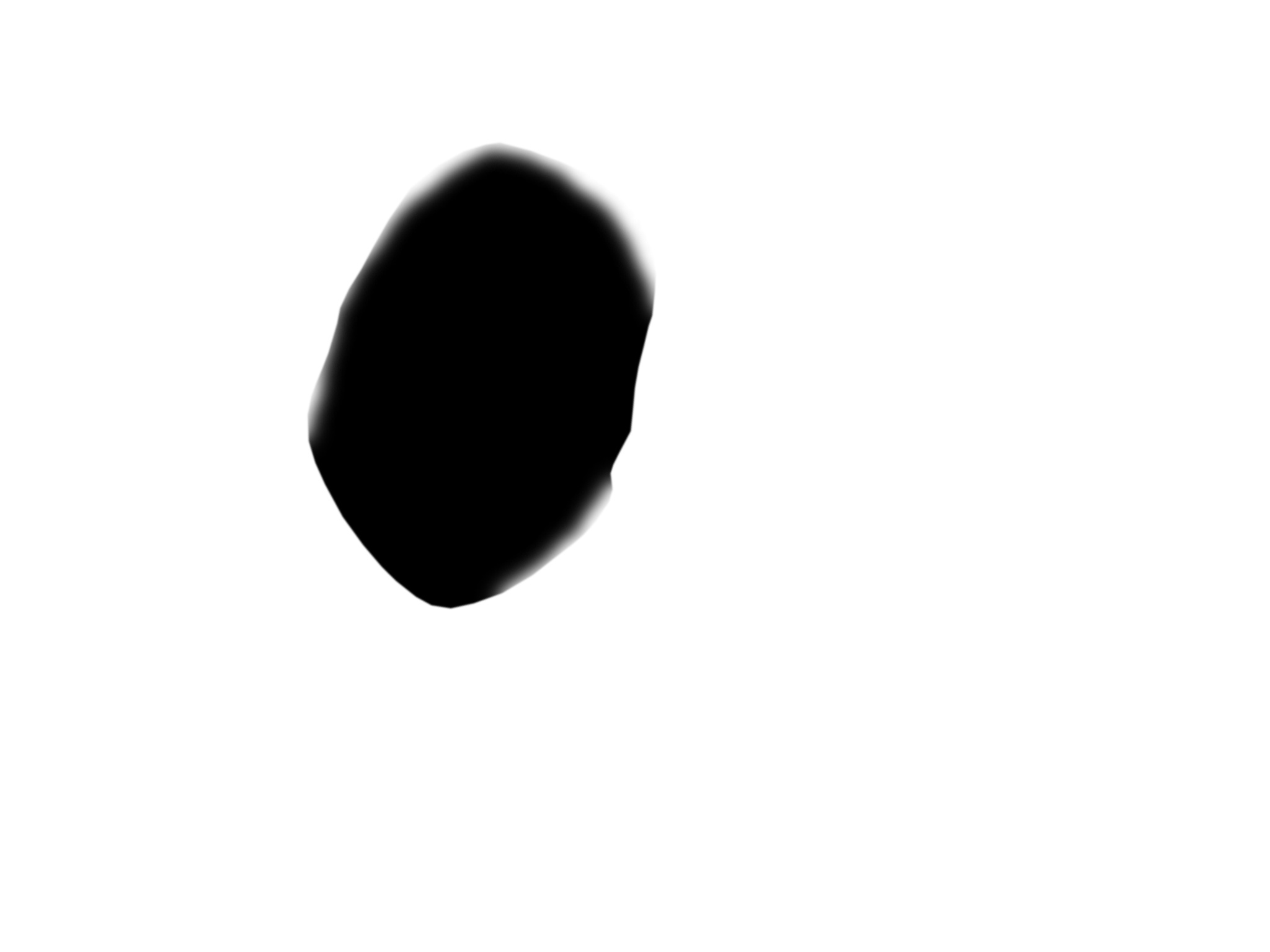}}}}
\end{subfloat}
\begin{subfloat}[FSG Heatmap\label{fsg}]
    {\fbox{\includegraphics[width=0.19\textwidth]{{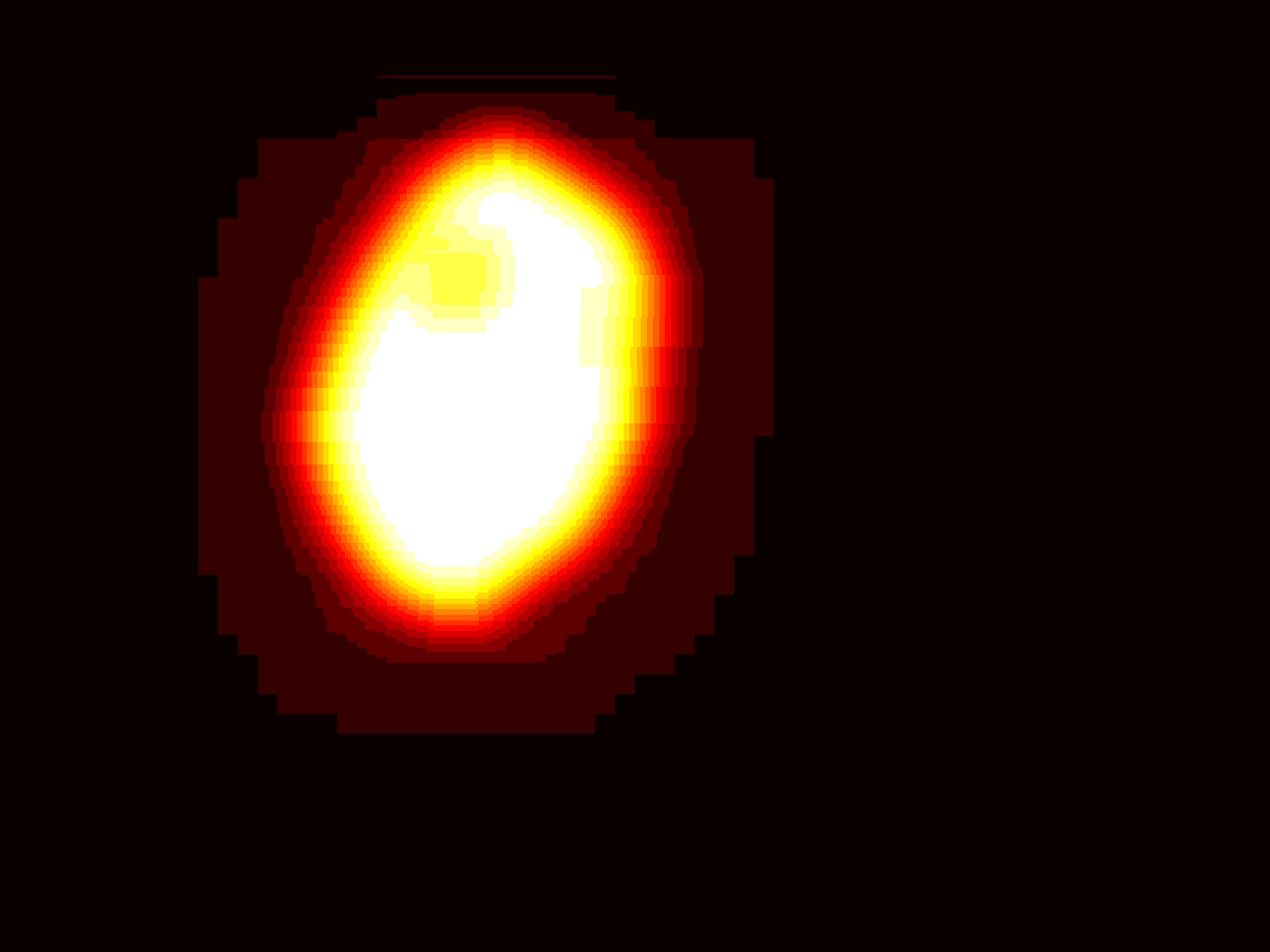}}}}
\end{subfloat}
\begin{subfloat}[EXIF-Net Heatmap\label{exifnet}]
    {\fbox{\includegraphics[width=0.19\textwidth]{{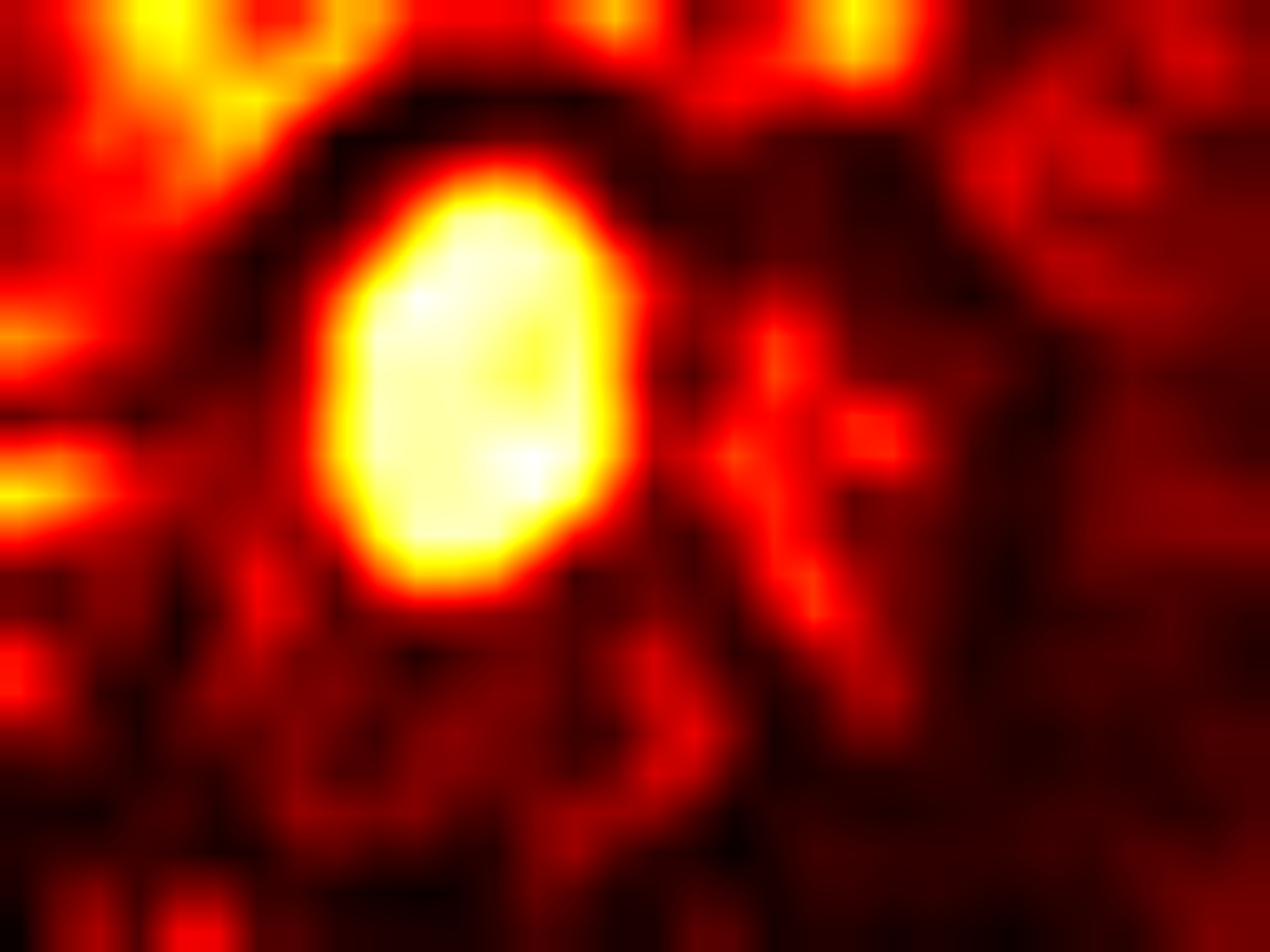}}}}
\end{subfloat}
\begin{subfloat}[Noiseprint Heatmap\label{np}]
    {\fbox{\includegraphics[width=0.19\textwidth]{{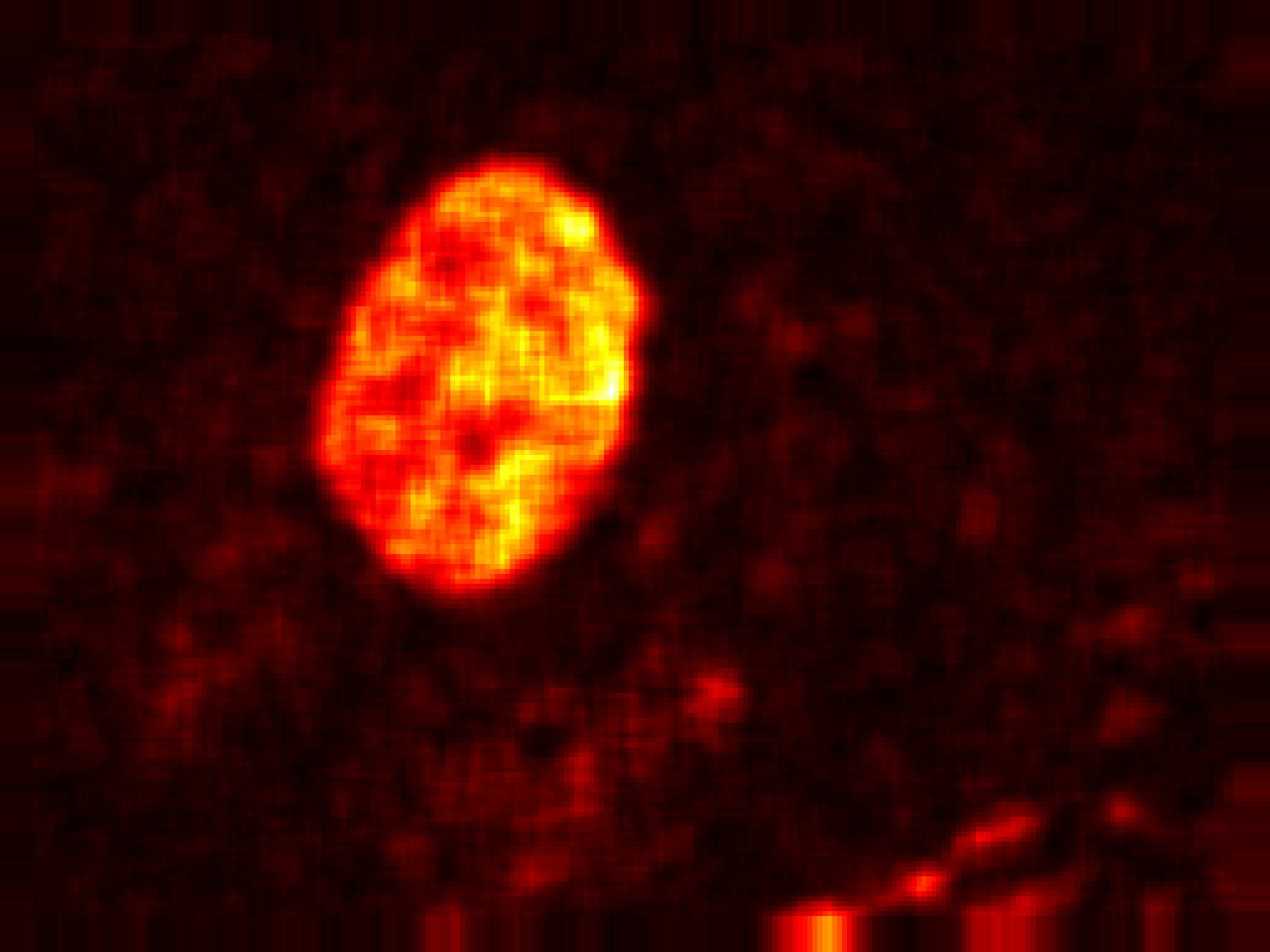}}}}
\end{subfloat}

\begin{subfloat}[Attacked Image\label{atkimage}]
    {\fbox{\includegraphics[width=0.19\textwidth]{{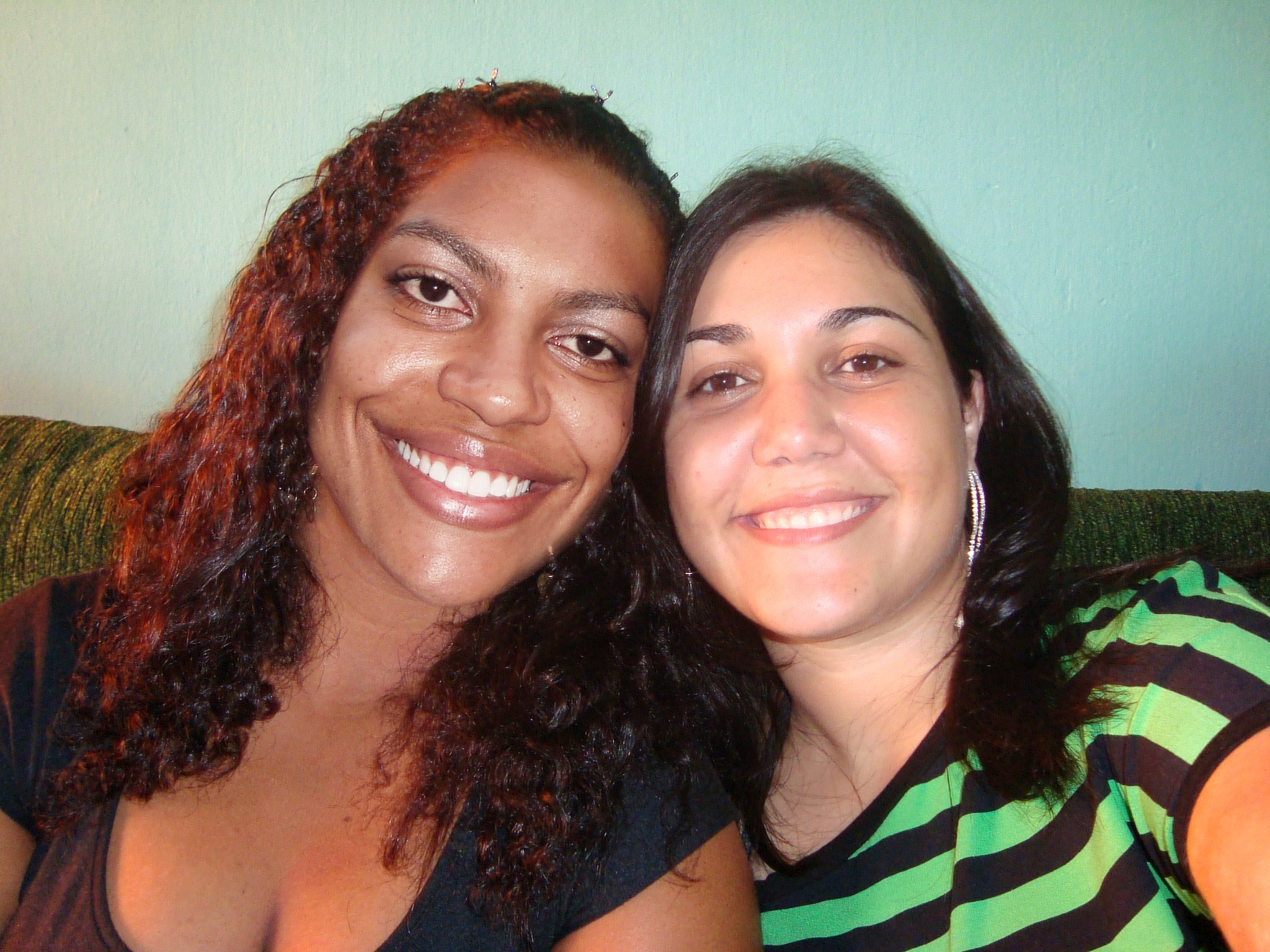}}}}
\end{subfloat}
\begin{subfloat}[Distortion\label{distortion}]
    {\fbox{\includegraphics[width=0.19\textwidth]{{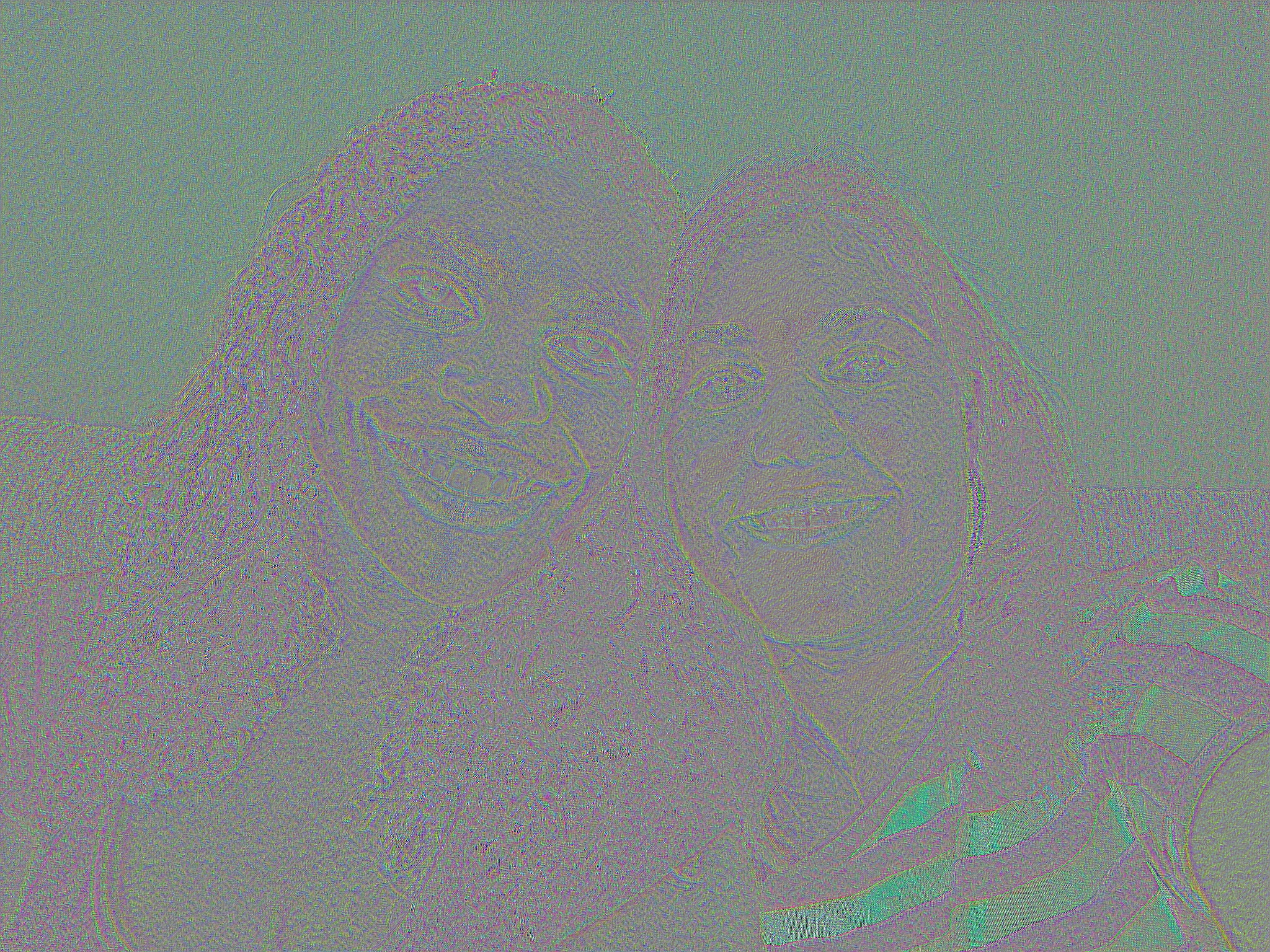}}}}
\end{subfloat}
\begin{subfloat}[FSG Heatmap\label{fsg-atk}]
    {\fbox{\includegraphics[width=0.19\textwidth]{{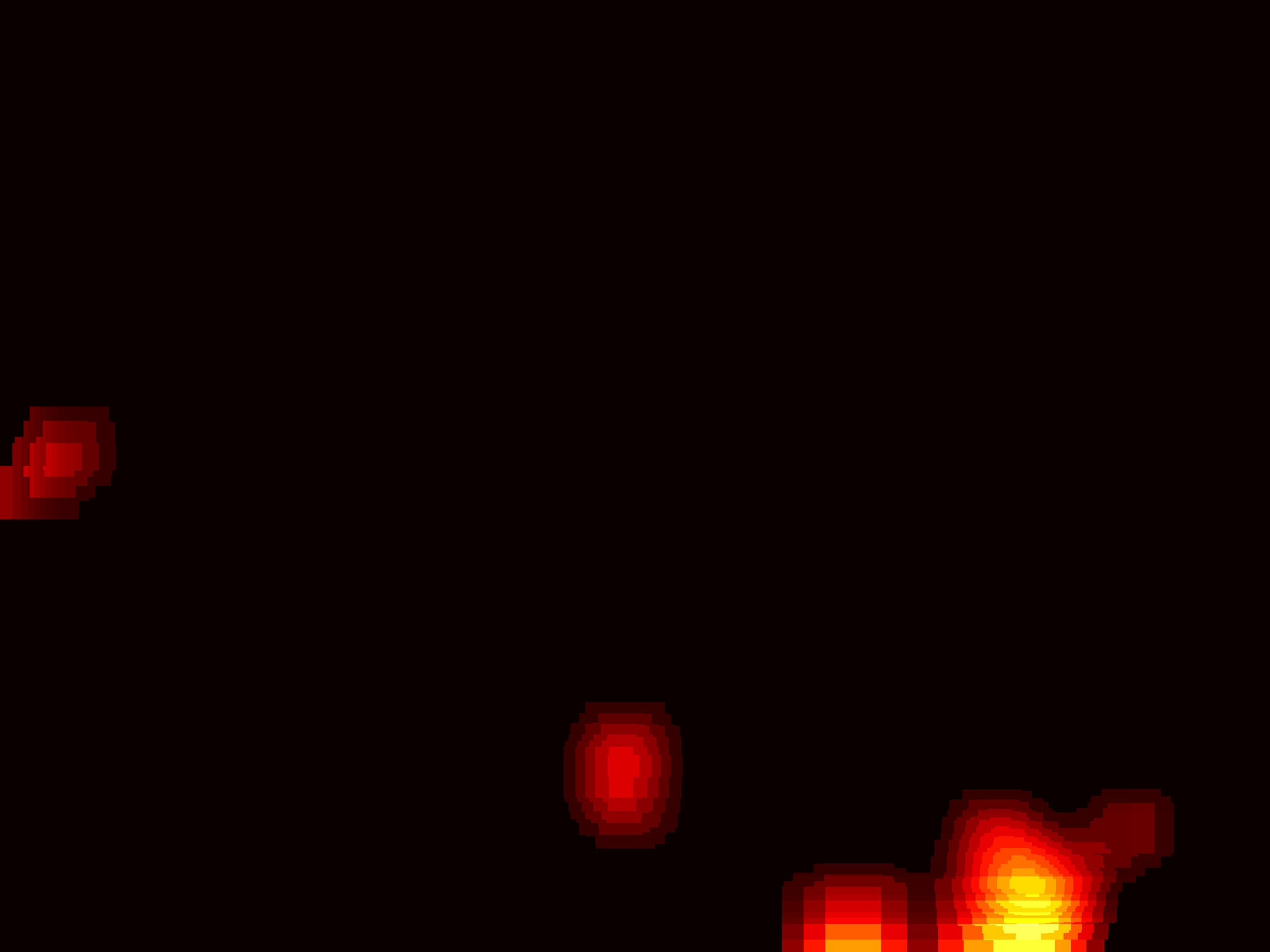}}}}
\end{subfloat}
\begin{subfloat}[EXIF-Net Heatmap\label{exifnet-atk}]
    {\fbox{\includegraphics[width=0.19\textwidth]{{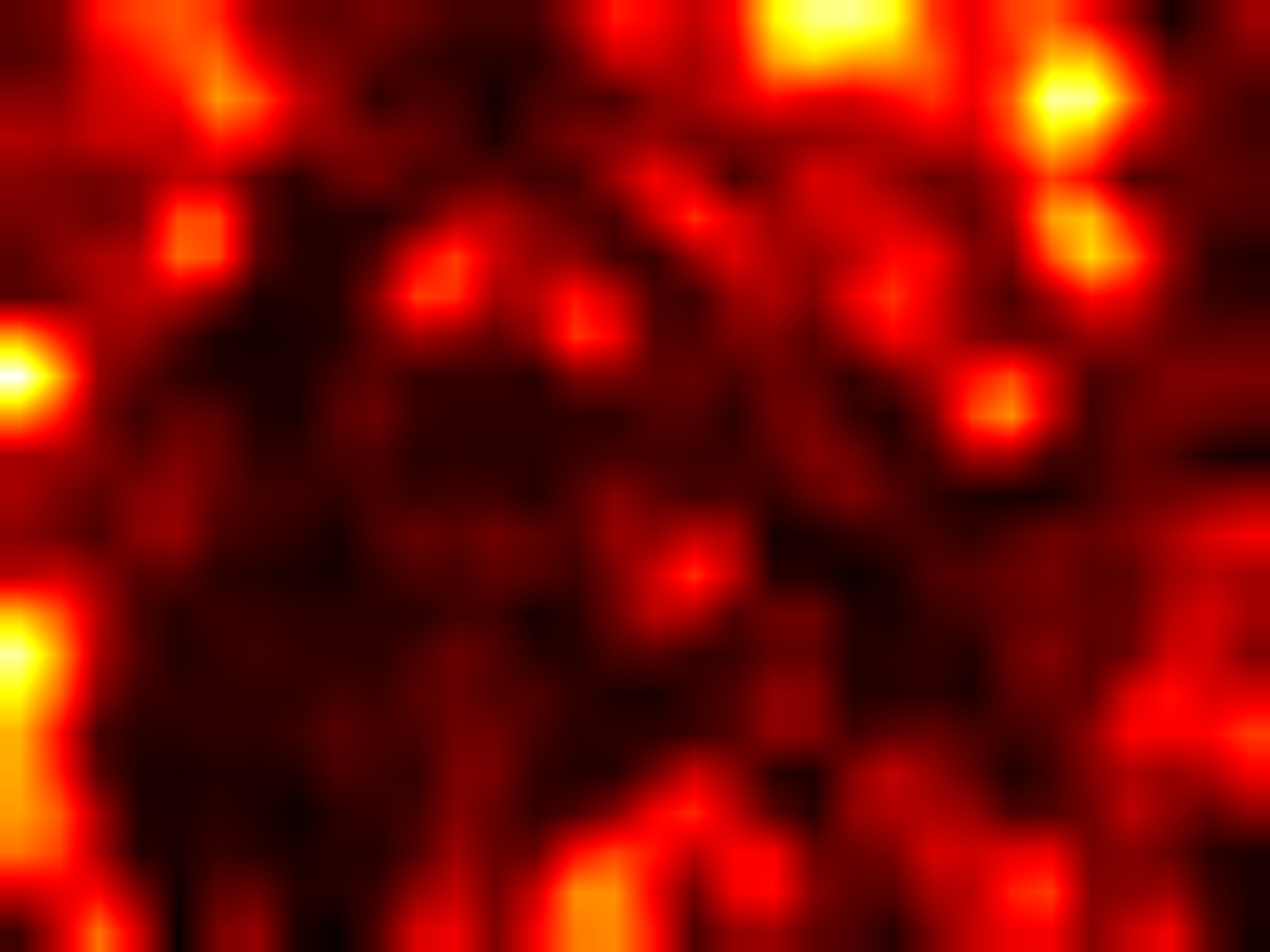}}}}
\end{subfloat}
\begin{subfloat}[Noiseprint Heatmap\label{np-atk}]
    {\fbox{\includegraphics[width=0.19\textwidth]{{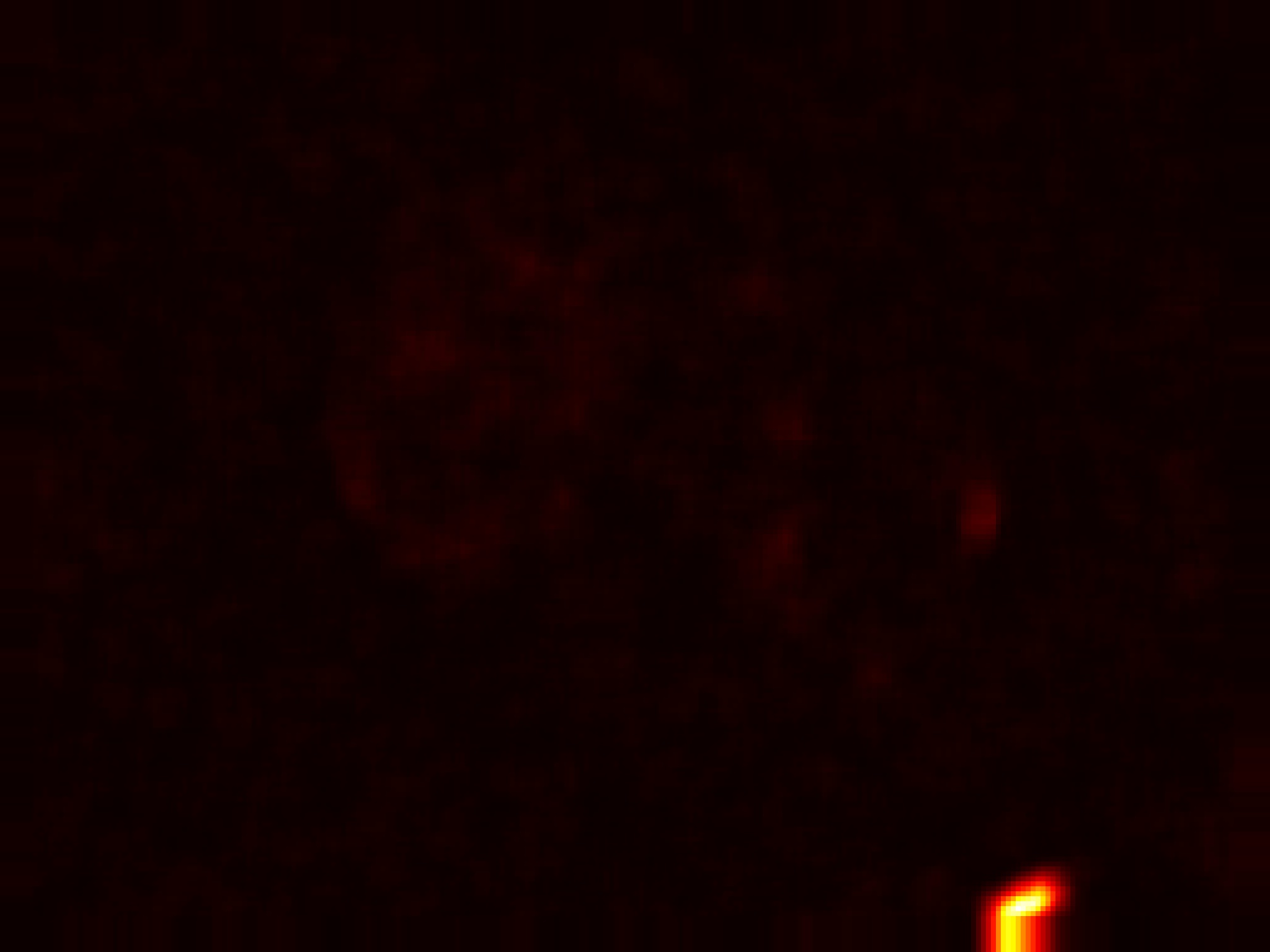}}}}
\end{subfloat}
\caption{Example from the Carvalho DSO-1 Database demonstrating how our attack can be used to fool the image splicing localizer. In this figure, the top row shows \protect\subref{spliceimage} the original spliced images, \protect\subref{grondtruth} the ground truth masks and the localization heatmaps on the original image produced by \protect\subref{fsg} FSG, \protect\subref{exifnet} EXIF-Net, and \protect\subref{np} Noiseprint. The bottom row shows \protect\subref{atkimage} the attacked image, \protect\subref{distortion} the perturbation introduced by our attack and the localization heatmaps on the attacked image produced by  \protect\subref{fsg-atk} FSG, \protect\subref{exifnet-atk} EXIF-Net, and \protect\subref{np-atk} Noiseprint. Note that we scale up the perturbation on the images 10 times for better visualization.}
\label{fig: CarvalhoExamples}
\end{figure*}

Our third experiment evaluated our attack's ability to fool image splicing localization algorithms.  In this experiment, we used the EXIF-Net, FSG, and Noiseprint algorithms to localize falsified content in attacked images.   The output of these algorithms were compared to ground truth masks to assess the performance of our attack.
We note that in these experiments, our attack was not explicitly trained against Noiseprint.  Successful attacks launched against Noiseprint provide an example of zero-knowledge attack (i.e. one in which we cannot train against the forensic algorithm in a white or black box setting) and provides information about the transferability of our attack.

\subhead{Dataset}
To conduct this experiment, we used the same databases as the previous experiment attacking splicing detection described in Section~\ref{sec:result_det}.
Because localization algorithms assume that an image is falsified a prior, 
 we only utilized the spliced images and their corresponding ground truth masks for evaluation. In total, we used 180 falsified images from the Columbia database, 100 falsified images from the Carvalho DSO-1 database, and 220 falsified images from the Korus databases. 

\subhead{Splice Localization Evaluation Metrics}
We quantified the performance of our attack on splice localization algorithms by comparing forensic localization algorithms' performance both before (i.e. baseline performance) and after our attack. 
We used the F-1 score and the Matthews correlation coefficient (MCC) score measured between a localizer's output and the ground truth mask to measure each localizer's performance.  These measures are commonly used in splicing localization literature~\cite{huh2018fighting,cozzolino2019noiseprint,mayer2020exposing}.  Significant drops in the F-1 and MCC score correspond to high attack performance.  These scores are defined as: 

\vspace{-0.9em}
{\footnotesize
\begin{align}
&F_{1} = \frac{2 \times Precision \times Recall}{Precision + Recall}\\
&MCC = \frac{TP \times TN - FP \times FN}{\sqrt{(TP + FP)(TP+FN)(TN+FP)(TN+FN)}}
\end{align} 
}%

Localization algorithms output a heat map identifying the potential spliced region.  As a result, the investigator must choose a threshold to compare this heatmap to in order to achieve a final prediction mask.  For this experiment, we always choose the threshold that yields the best localization scores for every individual spliced image for our evaluation.  This threshold choice is the ideal scenario for the investigator since it maximizes their localization performance, and the worst-case scenario for our attack. %

Furthermore, we note that all the localization methods utilize clustering algorithms to group image patches into two distinct regions to produce localization masks.   This occurs even for authentic images, which should in truth only contain one region.
Therefore, it is neither correct nor possible for an attack to make these localization algorithms predict no falsified region at all possible thresholds.
Therefore, we define an attack as successful when it can make the predicted region have a low correlation with the ground truth mask.  We assess this by observing the drop in the F-1 and MCC scores.  A localization mask that has low F-1 and MCC scores will lead an investigator to make incorrect decisions about what has been falsified in an image.

\subhead{Results}

\begin{table}[t]
\caption{Splicing localization performance, per image thresholds}
\label{LocalizationPerImg}
\small
\centering
\setlength{\tabcolsep}{1 mm}
\renewcommand{\arraystretch}{1.2}
\begin{tabular}{|c|c|c|c|c|c|c|c|}
\hline
 \multirow{2}{*}{Training $S_{n}$} &\multirow{2}{*}{\shortstack{Localization\\Method}} & \multicolumn{2}{c|}{Columbia} & \multicolumn{2}{c|}{DSO-1} & \multicolumn{2}{c|}{Korus}\\
 \cline{3-8}
 & & F1 & MCC & F1 & MCC & F1 & MCC\\
 \hline
 Baseline & \multirow{5}{*}{FSG} & 0.90 & 0.86 & 0.84 & 0.82 & 0.41 & 0.39 \\
 \cline{1-1}\cline{3-8} 
 CNN $C(\cdot)$ Only & & 0.80 & 0.73 & 0.49 & 0.41 & 0.34 & 0.33 \\
 \cline{1-1}\cline{3-8}  
 FSG & & 0.66 & 0.55 & 0.49 & 0.40 & 0.26 & 0.24 \\
 \cline{1-1}\cline{3-8} 
 EXIF-Net & & 0.82 & 0.77 & 0.65 & 0.60 & 0.37 & 0.37 \\
 \cline{1-1}\cline{3-8} 
 Both, No Init & & 0.71 & 0.60 & 0.49 & 0.41 & 0.25 & 0.23 \\
 \cline{1-1}\cline{3-8} 
 Both & & \bf{0.58} & \bf{0.44} & \bf{0.40} & \bf{0.30} & \bf{0.23} & \bf{0.22} \\
 \hline
 Baseline & \multirow{5}{*}{EXIF-Net} & 0.91 & 0.88 & 0.57 & 0.51 & 0.28 & 0.26 \\
 \cline{1-1}\cline{3-8} 
 CNN $C(\cdot)$ Only & & 0.90 & 0.87 & 0.50 & 0.43 & 0.27 & 0.26 \\
 \cline{1-1}\cline{3-8}  
 FSG & & 0.90 & 0.87 & 0.52 & 0.46 & 0.26 & 0.26 \\
 \cline{1-1}\cline{3-8} 
 EXIF-Net & & 0.73 & 0.65 & 0.38 & 0.26 & 0.23 & 0.21 \\
 \cline{1-1}\cline{3-8} 
 Both, No Init & & \bf{0.62} & \bf{0.51} & 0.38 & 0.28 & 0.23 & 0.21\\
 \cline{1-1}\cline{3-8} 
 Both & & 0.66 & 0.55 & \bf{0.36} & \bf{0.25} & \bf{0.22} & \bf{0.20} \\
 \cline{1-1}\hline
 Baseline & \multirow{5}{*}{Noiseprint} & 0.81 & 0.74 & 0.75 & 0.72 & 0.33 & 0.32 \\
 \cline{1-1}\cline{3-8} 
 CNN $C(\cdot)$ Only & & 0.78 & 0.71 & 0.68 & 0.63 & 0.31 & 0.29 \\
 \cline{1-1}\cline{3-8}  
 FSG & & 0.75 & 0.67 & 0.59 & 0.53 & 0.28 & 0.27 \\
 \cline{1-1}\cline{3-8} 
 EXIF-Net & & 0.76 & 0.68 & 0.64 & 0.59 & 0.34 & 0.32 \\
 \cline{1-1}\cline{3-8} 
 Both, No Init & & 0.75 & 0.68 & 0.50 & 0.42 & 0.27 & 0.26 \\
 \cline{1-1}\cline{3-8} 
 Both & & \bf{0.69} & \bf{0.59} & \bf{0.47} & \bf{0.37} & \bf{0.24} & \bf{0.22} \\
 \hline
\end{tabular}
\end{table}

To evaluate our attack's performance on splicing localization algorithms, we first launched our attack against the forged images in all three databases described above. Next we performed splicing localization using the FSG, EXIF-Net, and Noiseprint algorithms, and recorded the attack performance against each algorithm. Again, we evaluated our attack's performance using all four training strategies discussed previously 
as well as the Phase 1 only training strategy (i.e. `` (CNN $C(\cdot)$ Only'').

The localization results obtained from this experiment are shown in Table~\ref{LocalizationPerImg}. The lowest scores of each localization algorithms among all attack training strategies ( i.e  the highest attack performance) are marked in bold.

The results displayed in Table~\ref{LocalizationPerImg} show that after our attack, all three localization algorithms' performance dropped significantly. 
Our attack typically achieves the highest performance when it is trained using the `Both' strategy using Phase 1 initialization.  For example, using this training strategy our attack is able to drop FSG's F-1 score from 0.84 to 0.40 and its MCC score from 0.82 to 0.30 on the DS0-1 database.  Similarly, our attack is able to drop EXIF-Net's  F-1 score from 0.57 to 0.36 and its MCC score from 0.51 to 0.25 on the DS0-1 database.

We note that our attack can successfully transfer to attack Noiseprint when it is trained against both FSG and EXIF-Net using Phase 1 initialization.  This is significant, since we do not explicitly train against Noiseprint.  For example, our attack is able to drop Noiseprint's F-1 score from 0.75 to 0.47 and its MCC score from 0.72 to 0.37 on the DS0-1 database.  This demonstrates that it is possible for attack to transfer, i.e. successfully interfere with new localization algorithms that it was not trained against.  
We note that training only against a forensic CNN (i.e. only using Phase 1) as is done in prior GAN-based attacks is unsuccessful.

Visualizations showing sample results achieved by our attack are shown in Fig.~\ref{fig: CarvalhoExamples}.
These examples show localization results both before and after our attack is launched, along with the original images,  attacked images, and the ground truth splicing mask.  
From this figures, we can see that our attack can successfully hinder the performance of the FSG, EXIF-Net, and Noiseprint localization algorithms. 
Specifically, our attack can cause these localization algorithms to 
produce localization masks that are incorrect and uninformative.

Additionally, these examples show that no visually detectable traces are introduced into images by our attack.  We note that quantitative meausures of the distortion introduced by our attack are discussed in Section~\ref{sec:result_det} and displayed in Table~\ref{table: quality}.

While training our against multiple Siamese networks yielded the highest performance, we note that when our attack is trained only against the target $S_n$, our attack is able to cause large localization performance drops.
For example, on the Columbia database our attack trained only against FSG drops  FSG's F-1 score from 0.90 to 0.66 and its MCC score from 0.86 to 0.55.  Similarly, our attack trained against only EXIF-Net drops EXIF-Net's F-1 score from 0.91 to 0.73 and its MCC score from 0.88 to 0.65 on the Columbia database.   
This shows that targeted training attack is still successful, though training against a broader set of Siamese networks yields increased performance.

\subsection{Comparison with Other Attacks}
\label{sec:comparison}
There are several attacks that can be launched against neural networks using adversarial examples.  
Well known attacks such as the Fast Gradient Sign Method (FGSM)~\cite{goodfellow2015explain}, Projected Gradient Descent (PGD)~\cite{madry2018towards}, and Carlini-Wagner (CW)~\cite{carlini2017towards} directly modify an image's pixel values in order to fool a neural network.  
Other adversarial example attacks, such as the LOTS attack~\cite{Rozsa2017LOTS}, operate by modifying the intermediate feature maps of a neural network.  These attacks can be adapted to fool forensic neural networks~\cite{rozsa2020adversarial, Carlini2020CVPRW}.
Neural networks such as Siamese networks, however, typically only make up one part of forensic splicing detection and localization algorithms.  As a result, attacking these forensic algorithms is often more challenging than simply attacking their neural network components.

In this section, we compare our attack to each of the attacks mentioned above (FGSM PGD, CW, and LOTS).  Our experiments demonstrate that our attack outperforms each of these adversarial example based attacks.  
Additionally, we demonstrate important shortcomings of these attack approaches which our proposed attack addresses.

\subsubsection{Adversarial Example Attacks}

In our first experiment, we investigated the ability of ``end-to-end'' adversarial example attacks to fool splicing detection and localization algorithms.  End-to-end adversarial example attacks are the most common type of attacks.  They operate by directly modifying an image's pixel values in order to cause changes a neural network's output.  
In this experiment, we used the FGSM, PGD, and CW algorithms to launch attacks.  These algorithms were used to directly fool the %
$S_n$ that a %
forensic algorithm relies upon.  Specifically, $S_n$ produces a measure of the similarity of the traces in a pair of patches.  An end-to-end attack against $S_n$ operates by modifying the pixel values of an attacked patch so that $S_n$ believes they match the forensic traces in the other ``reference'' patch.
Fooling a splicing detection or localization algorithm, however, requires that all patches in an image appear to have a common set of traces.  To accomplish this, all patches in an image were attacked using a common ``reference'' patch in order to create consistent forensic traces throughout an image,

To implement these attacks, we used the Cleverhans library~\cite{papernot2018cleverhans}.  Attacks were directly launched against FSG and EXIF-Net.  
Noiseprint was not considered because the feature extractor doesn't directly rely upon a neural network classifier (or Siamese network).  As a result, it is not clear if Noiseprint can be attacked using an end-to-end adversarial example attack.
To choose the best parameters for each attack, we tried multiple parameter values and used those that yielded the best patch-wise attack performance. 
For PGD parameters, we set the maximum pixel distortion $\epsilon$ to be 3, the step size to be 1, and the maximum iteration to be 10 for both siamese networks. 
For FGSM, we chose a step size of 0.1 when  attacking both FSG and EXIF-Net. %
For the CW attack, we chose the step size to be 1e-2 for both Siamese networks, the loss ratio to be 800 for FSG and 30 for EXIF-Net. We also set the binary search steps to be 5. %

\begin{table}[t]
\caption{Splicing detection \& localization performance of different attacks, measured by F-1 and MCC score, using per image threshold}
\label{PGD}
\small
\centering
\setlength{\tabcolsep}{1 mm}
\renewcommand{\arraystretch}{1.2}
\begin{tabular}{|c|c|c|c|c|}
\hline
 \multirow{3}{*}{Attack} & \multirow{3}{*}{\shortstack{Evaluation\\Method}} & \multicolumn{3}{c|}{Columbia}\\
 \cline{3-5}
 & & Detection & \multicolumn{2}{c|}{Localization} \\
 \cline{3-5} 
 & & m-AP & F1 & MCC\\
 \hline
 Baseline     & \multirow{8}{*}{EXIF-Net}  & 0.98 & 0.91 & 0.88 \\
 \cline{1-1}\cline{3-5}
 PGD-FSG      & & \bf{0.68} & 0.87 & 0.83 \\
 \cline{1-1}\cline{3-5}
 PGD-EXIF-Net  & & 0.90 & \bf{0.71} & \bf{0.62} \\
 \cline{1-1}\cline{3-5}
 CW-FSG       & & 0.98 & 0.91 & 0.88 \\
 \cline{1-1}\cline{3-5} 
 CW-EXIF-Net   & & 0.98 & 0.91 & 0.88 \\
 \cline{1-1}\cline{3-5} 
 FGSM-FSG     & & 0.98 & 0.90 & 0.88 \\
 \cline{1-1}\cline{3-5} 
 FGSM-EXIF-Net & & 0.96 & 0.85 & 0.81 \\
 \cline{1-1}\cline{3-5}
 Our proposed & & \bf{0.38} & \bf{0.66} & \bf{0.55} \\
 \hline
 Baseline & \multirow{8}{*}{FSG} & 0.94 & 0.90 & 0.86 \\
 \cline{1-1}\cline{3-5} 
 PGD-FSG & & \bf{0.71} & \bf{0.58} & \bf{0.41} \\
 \cline{1-1}\cline{3-5}
 PGD-EXIF-Net & & 0.77 & 0.80 & 0.73 \\
 \cline{1-1}\cline{3-5}
 CW-FSG & & 0.92 & 0.88 & 0.83 \\
 \cline{1-1}\cline{3-5}
 CW-EXIF-Net & & 0.94 & 0.90 & 0.86 \\
 \cline{1-1}\cline{3-5}
 FGSM-FSG & & 0.85 & 0.78 & 0.72 \\
 \cline{1-1}\cline{3-5}
 FGSM-EXIF-Net & & 0.93 & 0.88 & 0.85 \\
 \cline{1-1}\cline{3-5}
 Our proposed & & \bf{0.46} & \bf{0.58} & \bf{0.44} \\
 \hline
\end{tabular}
\end{table}

In Table.~\ref{PGD}, we show the  performance of the FGSM, CW, and PGD attacks as well as our proposed attack. The leftmost column displays the baseline performance of each forensic algorithm (i.e. without any attacks), as well as the forensic algorithm's performance under  PGD, CW, and FGSM attacks against both Siamese networks (FSG or EXIF-Net), and our proposed both-attack. We mark the best performance among all adversarial example attacks as well as our proposed attack's performance  in bold. 

These results show that our proposed attack outperforms end-to-end adversarial example attacks.   
Specifically, end-to-end attacks typically do not successfully reduce either forensic detector's m-AP to below the threshold of a random guess (i.e.~m-AP$\,\,\leq 0.5$).  
For example, the adversarial example attacks encountered significant difficulty dropping the m-AP score of EXIF-Net below 0.90.  This means most forgeries could still be very easily detected by EXIF-Net even after adversarial example attacks were launched. 
The only adversarial example attack that could do this was PGD trained against FSG, which lowered the m-AP score to 0.68 (notably, still above the random guess threshold).  By contrast, our proposed attack dropped the m-AP score from 0.98 to 0.38.  This means our proposed attack dropped EXIF-Net's detection performance to worse than a random guess, while the end-to-end adversarial example attacks could not.

Similarly, our attack typically outperforms these end-to-end adversarial example attacks by a significant margin.  From Table~\ref{PGD}, we can see that when attacking FSG, most adversarial example attacks could not drop the F1 score below 0.78 and the MCC score below 0.72. By contrast, our attack was able to drop the F1 score to 0.58 and the MCC score to 0.44.  

The only adversarial example attack that could achieve comparable performance was PGD trained against FSG.  We note, however, that this attack cannot drop the detector's performance below an m-AP of 0.71.  As a result, the attack will be easily detected.  Because of this, an analyst will decide that an image is untrustworthy, thus making an unsuccessful attack.  
Furthermore, when end-to-end attacks are able to produce low localization scores (for example PGD), they tend to do this by making most patches in an image appear inconsistent on a random basis (i.e. a significant number of the patches appear to come from a different source).  While this produces a low localization score, it simultaneously produces a high detection score.  As a result, these images will be quickly flagged as fake.  This results in an overall unsuccessful attack.

\subsubsection{LOTS Attack}

\begin{table}[t]
\caption{Splicing localization performance of LOTS-based attack and our proposed attack, measured by F-1 score and MCC score, using per image thresholds}
\label{ALLLOTS}
\small
\centering
\setlength{\tabcolsep}{1 mm}
\renewcommand{\arraystretch}{1.2}
\begin{tabular}{|c|c|c|c|c|c|c|c|}
\hline
 \multirow{2}{*}{Attack} & Evaluation & \multicolumn{2}{c|}{Columbia} & \multicolumn{2}{c|}{DSO-1} & \multicolumn{2}{c|}{Korus RT}\\
 \cline{3-8}
 & Method & F1 & MCC & F1 & MCC & F1 & MCC\\
 \hline
 Baseline & \multirow{3}{*}{FSG} & 0.90 & 0.86 & 0.84 & 0.82 & 0.41 & 0.39 \\
 \cline{1-1}\cline{3-8}
 LOTS Best & & 0.82 & 0.77 & 0.66 & 0.63 & 0.38 & 0.37 \\ %
 \cline{1-1}\cline{3-8}
 Our proposed & & \bf{0.58} & \bf{0.44} & \bf{0.40} & \bf{0.30} & \bf{0.23} & \bf{0.22} \\ 
 \hline
 Baseline & \multirow{3}{*}{EXIF-Net} & 0.91 & 0.88 & 0.57 & 0.51 & 0.28 & 0.26 \\
 \cline{1-1}\cline{3-8}
 LOTS Best & & 0.68 & 0.57 & 0.37 & 0.27 & 0.26 & 0.24 \\
 \cline{1-1}\cline{3-8}
 Our proposed &  & \bf{0.66} & \bf{0.55} & \bf{0.36} & \bf{0.25} & \bf{0.22} & \bf{0.20} \\
 \hline
 Baseline & \multirow{3}{*}{Noiseprint} & 0.81 & 0.74 & 0.75 & 0.72 & 0.33 & 0.32 \\
 \cline{1-1}\cline{3-8}
 LOTS Best & & 0.73 & 0.64 & 0.66 & 0.62 & 0.35 & 0.33 \\
 \cline{1-1}\cline{3-8}
 Our proposed &  & \bf{0.69} & \bf{0.59} & \bf{0.47} & \bf{0.37} & \bf{0.24} & \bf{0.22} \\
 \hline
\end{tabular}
\end{table}

\begin{table}[t]
\caption{Splicing localization performance of the EXIF-Net, under different analysis block grids, measured by F-1 score and MCC score, using per image thresholds}
\label{LOTSandBoth}
\small
\centering
\setlength{\tabcolsep}{1 mm}
\renewcommand{\arraystretch}{1.2}
\begin{tabular}{|c|c|c|c|c|c|c|c|}
\hline
 \multirow{2}{*}{Attack} & Sampling & \multicolumn{2}{c|}{Columbia} & \multicolumn{2}{c|}{DSO-1} & \multicolumn{2}{c|}{Korus RT}\\
 \cline{3-8}
 & Patches & F1 & MCC & F1 & MCC & F1 & MCC\\
 \hline
 \multirow{4}{*}{LOTS-EXIF-Net} & 23 & 0.70 & 0.60 & 0.39 & 0.29 & 0.24 & 0.22 \\
 \cline{2-8} & 30(default) & \bf{0.68} & \bf{0.57} & \bf{0.37} & \bf{0.27} & \bf{0.26} & \bf{0.24} \\
 \cline{2-8} & 37 & 0.71 & 0.61 & 0.43 & 0.34 & 0.27 & 0.25 \\
 \cline{2-8} & 45 & 0.71 & 0.61 & 0.42 & 0.34 & 0.27 & 0.26 \\
 \hline
 \multirow{4}{*}{Our proposed} & 23 & 0.65 & 0.53 & 0.35 & 0.23 & 0.21 & 0.18 \\
 \cline{2-8} & 30(default) & \bf{0.66} & \bf{0.55} & \bf{0.36} & \bf{0.25} & \bf{0.22} & \bf{0.20} \\
 \cline{2-8} & 37 & 0.65 & 0.54 & 0.36 & 0.25 & 0.22 & 0.19 \\
 \cline{2-8} & 45 & 0.66 & 0.55 & 0.37 & 0.26 & 0.22 & 0.19 \\
 \hline
\end{tabular}
\end{table}

In our next experiment, we investigated the performance of the LOTS-based attack~\cite{rozsa2020adversarial}.  
LOTS operates by attacking the feature maps of a neural network to synthesize pixel values that fool it.  Since LOTS attacks these intermediate feature maps, it is not an ``end-to-end'' attack. 
To anti-forensically attack an image, LOTS launches a series of attacks on each image patch analyzed by a forensic detection or localization algorithm.  We refer to the specific locations of the patches used by the forensic algorithm as the analysis block grid.  LOTS modifies each patch in the analysis block grid so that their feature maps match the average feature map of authentic patches.

We first compared the performance of our attack to the LOTS-based attack using the ``best-case scenario'' for LOTS (i.e. the attacker perfectly knows the analysis blocking grid and splicing mask).  This provides an upper bound on the performance of LOTS in comparison to our proposed attack.  In later experiments, we show that if side information is not available to the LOTS attack, its performance can drop.  In this experiment, we compared the splicing localization performance of each forensic algorithm (FSG, EXIF-Net, and Noiseprint) both before an attack (i.e. Baseline) and after each attack.  We note here that our attack was not explicitly trained against the Noiseprint localization algorithm.

Table~\ref{ALLLOTS} shows the results of this experiment, with the best attack performance against each localization algorithm marked in bold.  These results show that our proposed attack outperforms the best-case LOTS attack against all localization algorithms on all datasets.  In several cases, our attack significantly outperformed the LOTS attack.  
For example, when attacking FSG our attack dropped the F1 score from 0.75 to 0.47 and the MCC score from 0.72 to 0.37 on the DSO-1 dataset.  By contrast, the LOTS attack only dropped the F1 score to 0.66 and the MCC score to 0.62.  
This corresponds to our attack producing a drops in the F1 and MCC scores that are  255\% greater and 350\% greater respectively  than the drops induced by LOTS.

Furthermore, we note that our attack was not trained against the Noiseprint algorithm in these experiments.  The significant drop in Noiseprint's localization performance seen in Table.~\ref{ALLLOTS} demonstrates that our attack can transfer to other localization algorithms that it was not explicitly trained against.  %

We note that  LOTS  assumes that the attacker exactly knows the analysis block grid used by the forensic algorithm under attack, however,  this is not realistic.  In reality, the forensic algorithm may overlap analysis blocks by an amount that is not known to the attacker.  For example, by default EXIF-Net  divides the longest dimension of an image into 30 overlapping patches.  A user, however, can specify the number of ``sampling patches'',  i.e. the number of overlapping analysis blocks along the longest image dimension.  As a result, the analysis block grid used by the attacker may not align with the one used by the forensic algorithm.  This can potentially drop the performance of the LOTS attack.

To demonstrate this phenomenon, we ran an experiment where we allowed the analysis block grid to be misaligned with the attack block grid.  Results of this are displayed in Table.~\ref{LOTSandBoth}, which shows the localization performance of our both LOTS and our proposed algorithm using different block grids.    The ``Sampling Patches" are the number of patches along the longest dimension of the full-size image used by the analysis block grid. 

From this table we can see that while our attack's performance holds nearly constant as the attack block grid and analysis block grid are misaligned, the performance of the LOTS attack varies.  For example, when evaluating LOTS on the DSO-1 dataset the F1 score increases from 0.37 to 0.42 and the MCC score increases  from 0.27 to 0.34 when the number of sampling patches increases to 45.  Though these numbers do not appear to be large, they correspond to a $14\%$ increase in the F1 score and a $26\%$ increase in the MCC score.  By contrast, our proposed attack only sees increases of 0.01 in both the F1 and MCC scores, corresponding to increase of $3\%$ and $4\%$ respectively.

\begin{figure*}[t]
\centering
\setlength{\fboxsep}{0pt}
\hspace{0.19\textwidth}
\begin{subfloat}[Pristine Image\label{pristine}]
    {\fbox{\includegraphics[width=0.19\textwidth]{{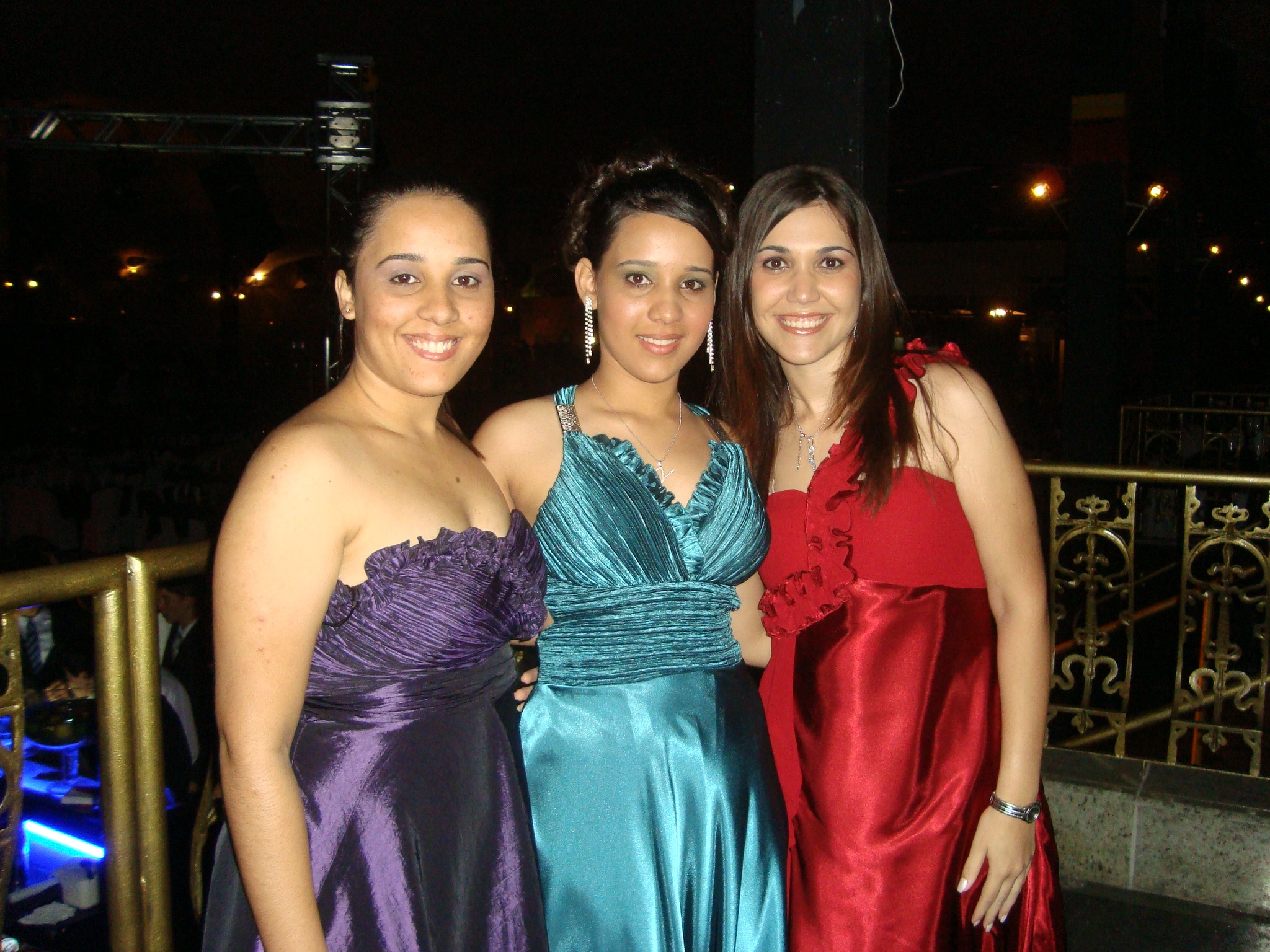}}}}
\end{subfloat}
\begin{subfloat}[FSG Heatmap\label{fsg1}]
    {\fbox{\includegraphics[width=0.19\textwidth]{{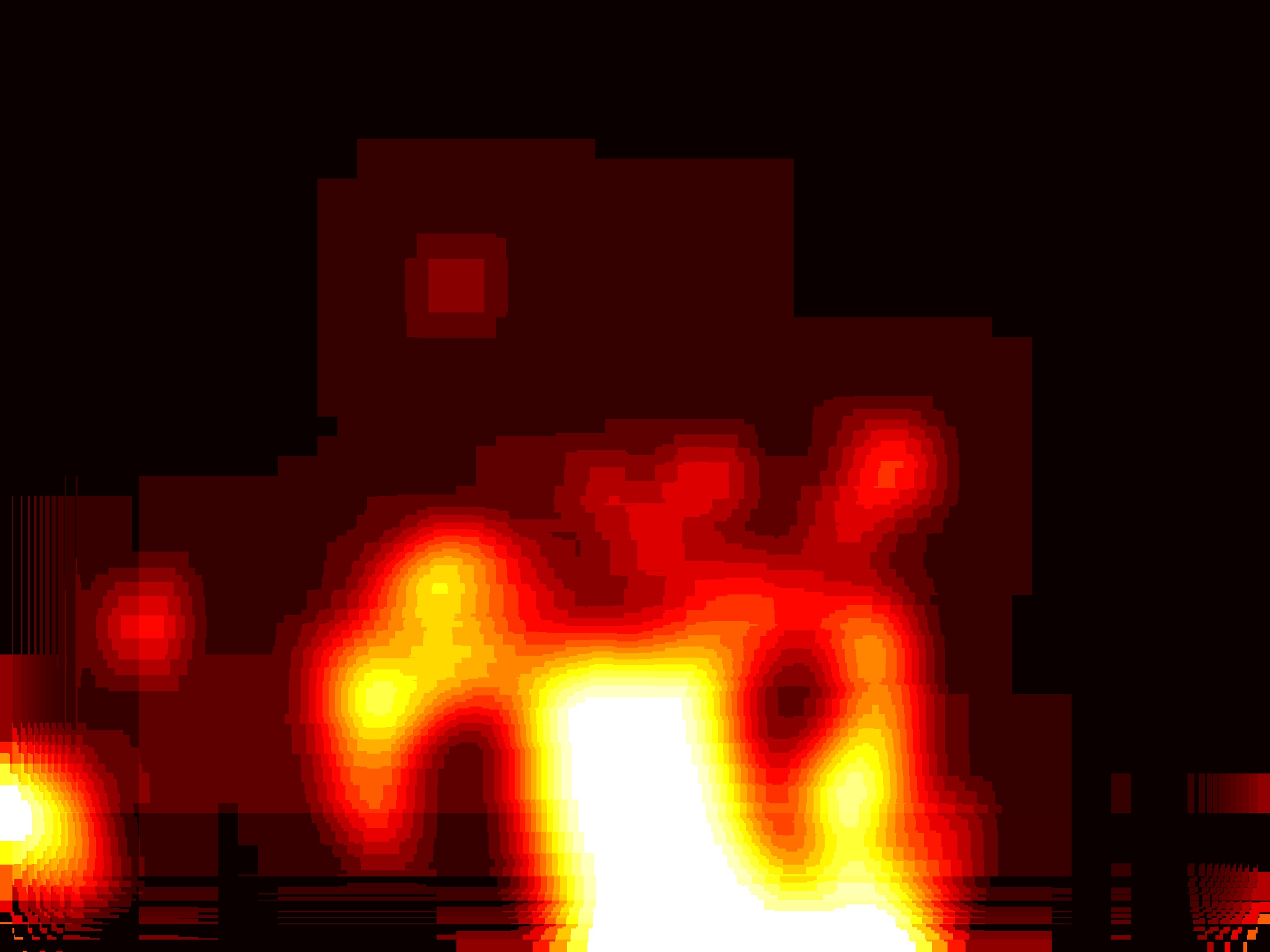}}}}
\end{subfloat}
\begin{subfloat}[EXIF-Net Heatmap\label{exifnet1}]
    {\fbox{\includegraphics[width=0.19\textwidth]{{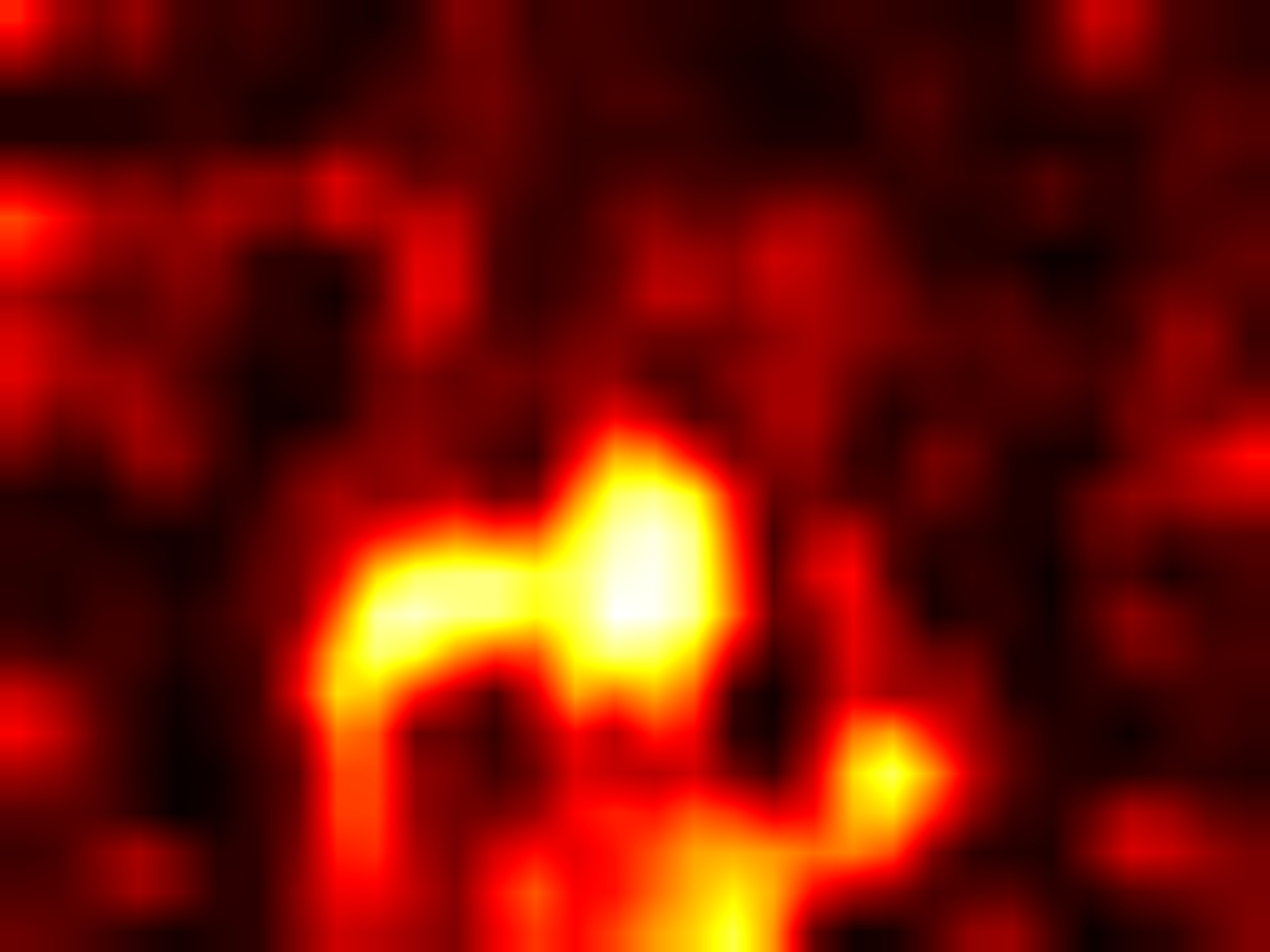}}}} 
\end{subfloat}
\begin{subfloat}[Noiseprint Heatmap\label{np1}]
    {\fbox{\includegraphics[width=0.19\textwidth]{{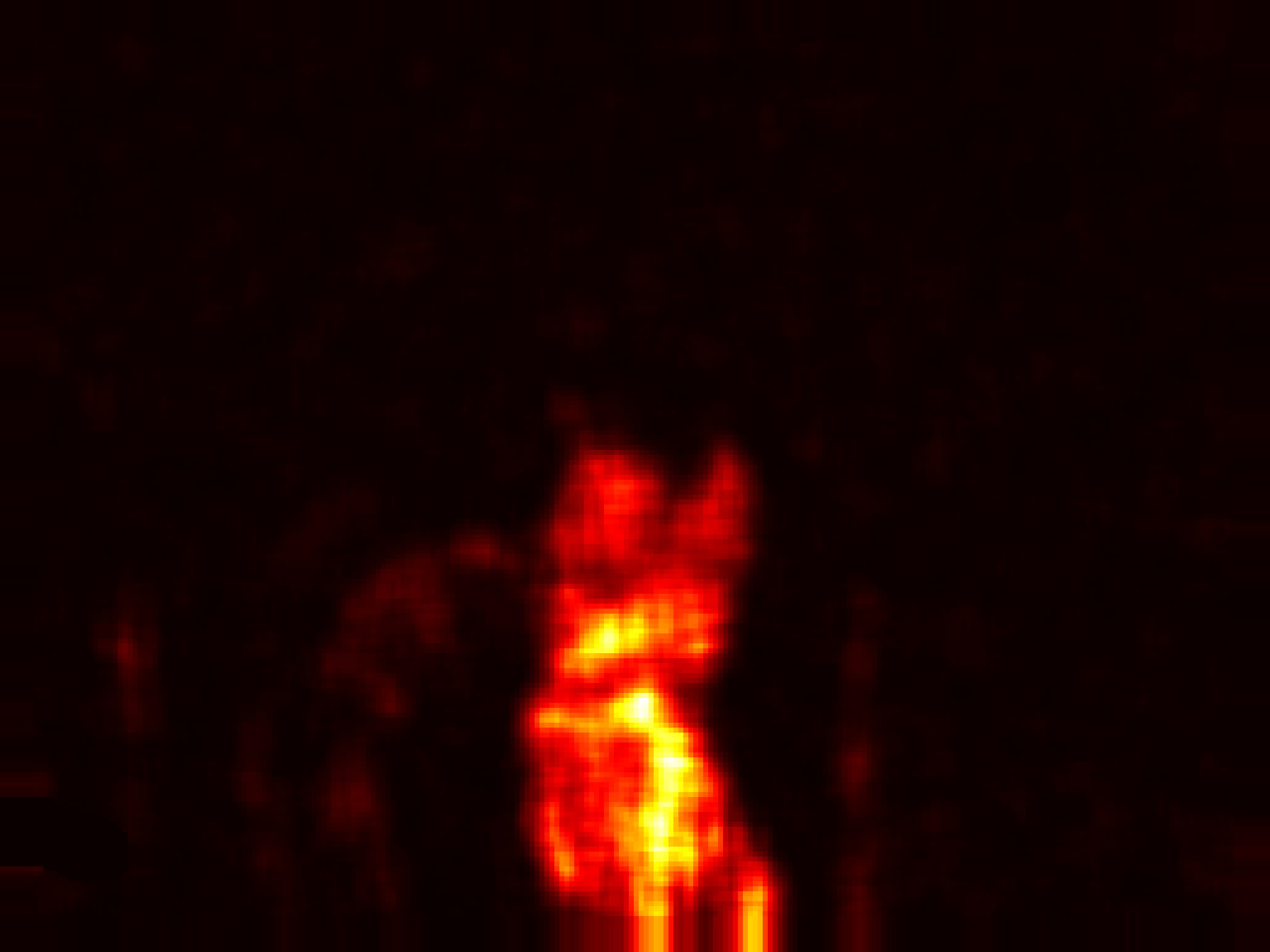}}}} 
\end{subfloat}
\begin{subfloat}[Attack Mask\label{mask1}]
    {\fbox{\includegraphics[width=0.19\textwidth]{{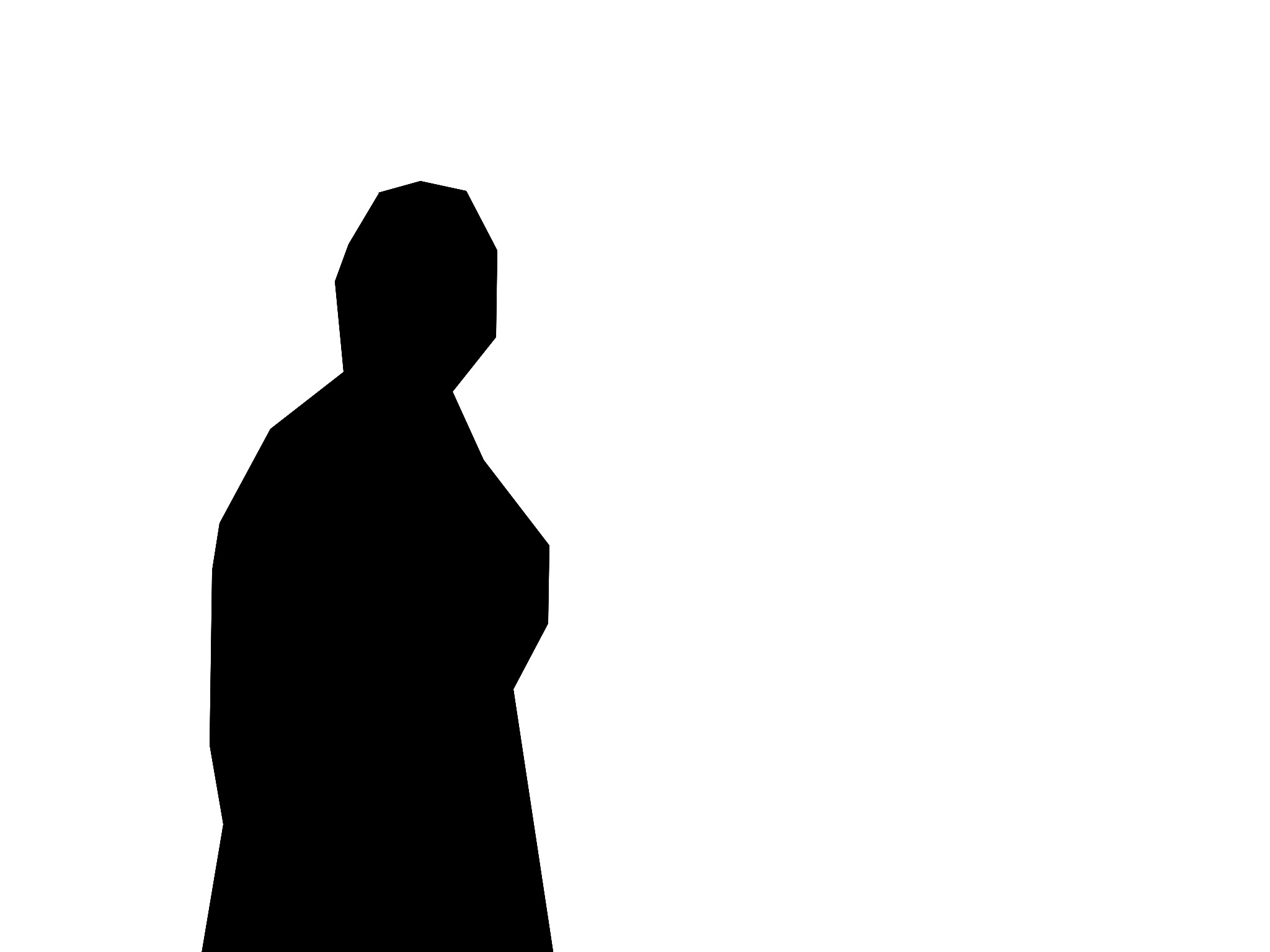}}}}
\end{subfloat}
\begin{subfloat}[Attacked Image\label{atkimage1}]
    {\fbox{\includegraphics[width=0.19\textwidth]{{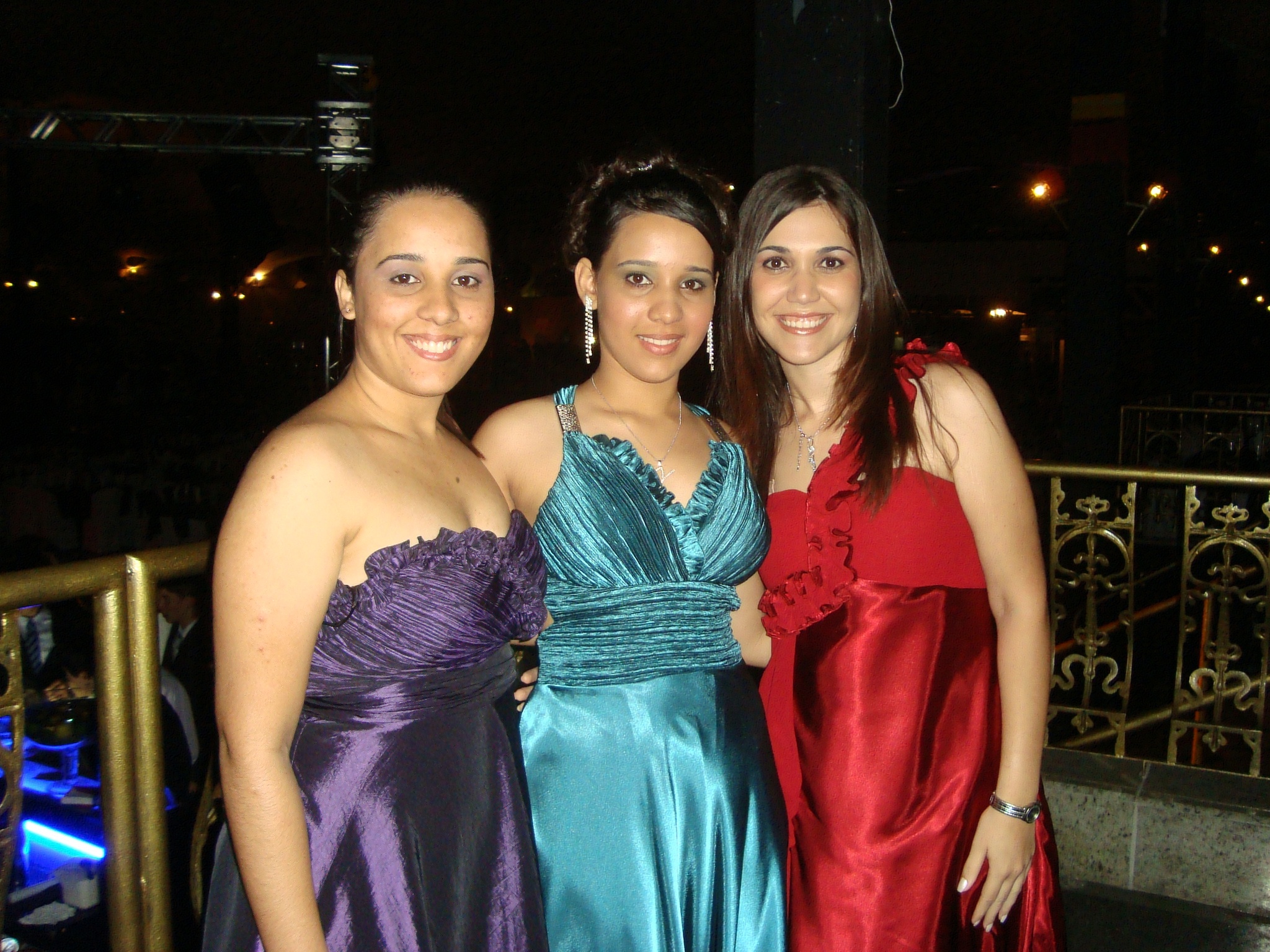}}}}
\end{subfloat}
\begin{subfloat}[FSG Heatmap\label{fsg1-atk}]
    {\fbox{\includegraphics[width=0.19\textwidth]{{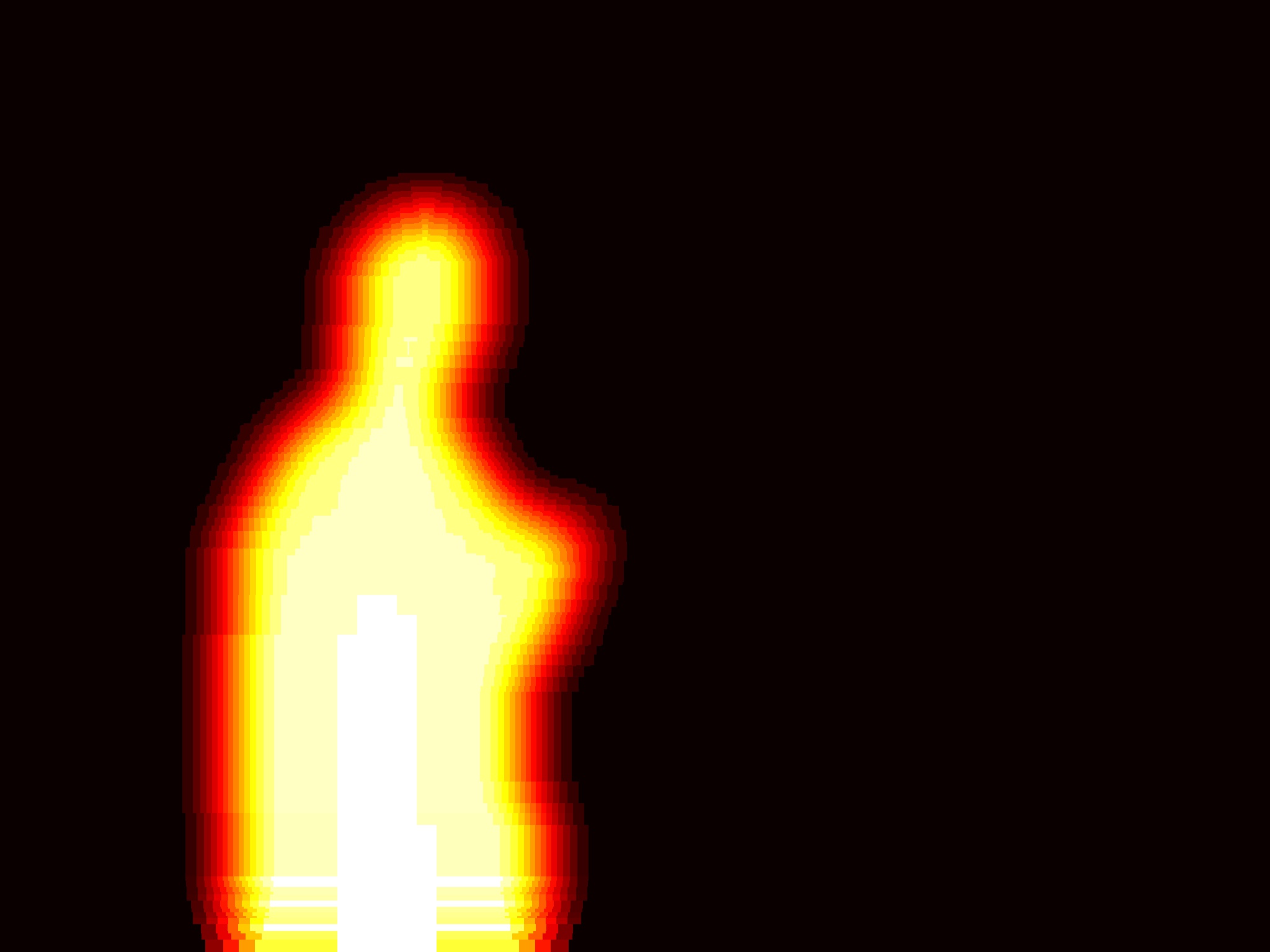}}}}
\end{subfloat}
\begin{subfloat}[EXIF-Net Heatmap\label{exifnet1-atk}]
    {\fbox{\includegraphics[width=0.19\textwidth]{{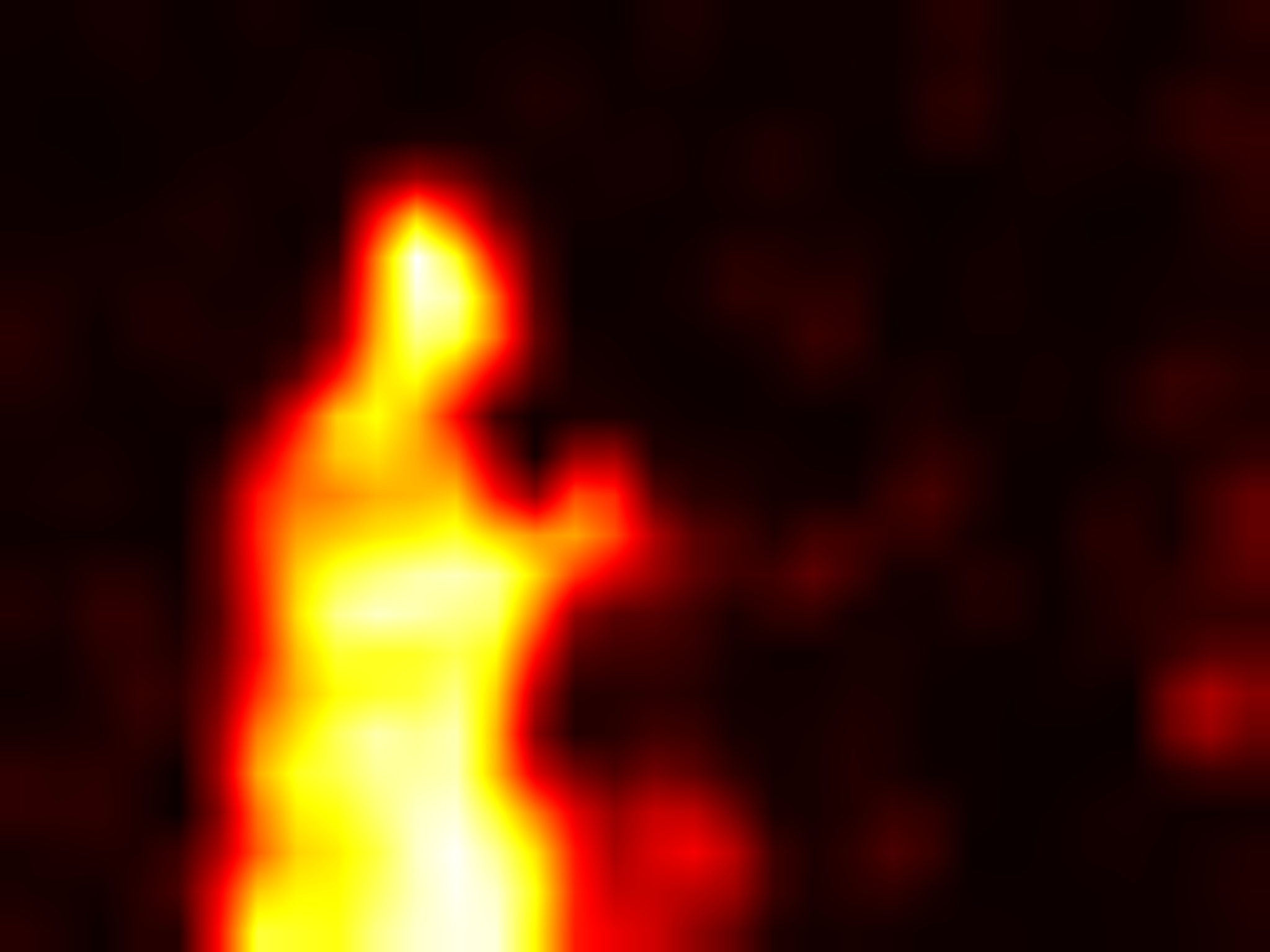}}}} 
\end{subfloat}
\begin{subfloat}[Noiseprint Heatmap\label{np1-atk}]
    {\fbox{\includegraphics[width=0.19\textwidth]{{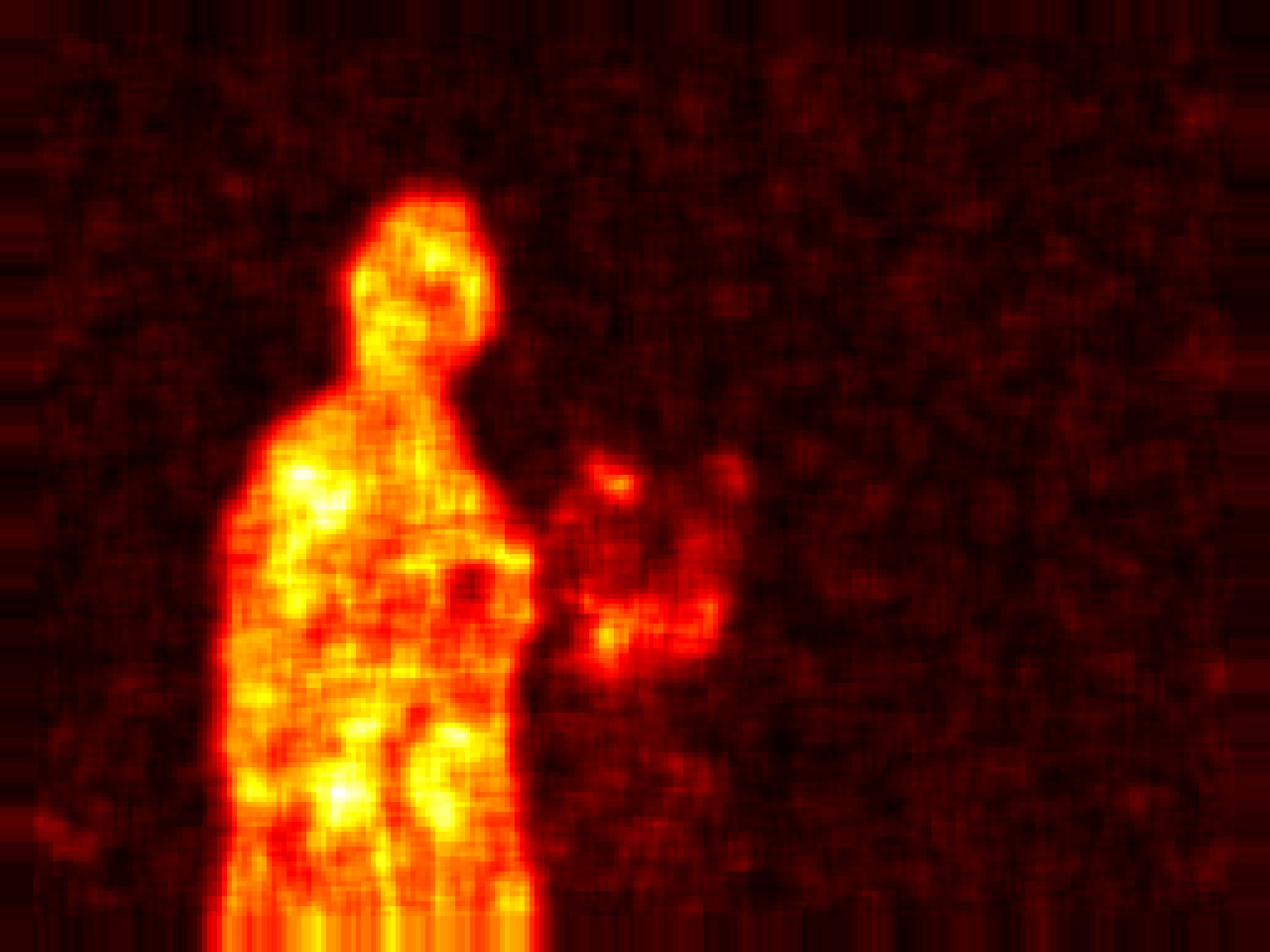}}}}
\end{subfloat}
\caption{Example demonstrating how our attack can be used to make an authentic image appear fake.  
In this figure, the top row shows \protect\subref{pristine} the original unaltered image and the localization heatmaps produced by \protect\subref{fsg1} Forensic Similarity Graph, \protect\subref{exifnet1} EXIF-Net, and \protect\subref{np1} Noiseprint.  The bottom row shows \protect\subref{mask1} the attack's target splicing mask, \protect\subref{atkimage1} the attacked image, and localization masks produced by analyzing the attacked image usinsg \protect\subref{fsg1-atk} Forensic Similarity Graph, \protect\subref{exifnet1-atk} EXIF-Net, and \protect\subref{np1-atk} Noiseprint.}
\label{fig:falsify_pristine}
\end{figure*}

Additionally, the LOTS-based attack assumes that the attacker has a mask indicating the falsified image region.  As a result, the attacker can use this information to identify unaltered image patches to form the target feature map.  However, this assumption is not always realistic.  The attacker might attack fake images gathered from the wild in which they do not accurately know the mask (e.g. as part of a misinformation campaign) or the spliced region may be so irregular that such patches do not exist.   %

\begin{table}[t]
\caption{Splicing Localization performance of LOTS-based attack on EXIF-Net, with and without masks, under different sampling method, measured by F-1 and MCC score, using per image threshold}
\label{No-Mask}
\small
\centering
\setlength{\tabcolsep}{1 mm}
\renewcommand{\arraystretch}{1.2}
\begin{tabular}{|c|c|c|c|c|c|}
\hline
 \multirow{2}{*}{Attack} & \multirow{2}{*}{\shortstack{Sampling \\ Patches}} & \multicolumn{2}{c|}{Columbia} & \multicolumn{2}{c|}{DSO-1}\\
 \cline{3-6}
 & & F1 & MCC & F1 & MCC \\
 \hline
 \multirow{4}{*}{LOTS-EXIF-Net} & 23 & 0.70 & 0.60 & 0.39 & 0.29 \\
 \cline{2-6} & 30(default) & \bf{0.68} & \bf{0.57} & \bf{0.37} & \bf{0.27} \\
 \cline{2-6} & 37 & 0.71 & 0.61 & 0.43 & 0.34 \\
 \cline{2-6} & 45 & 0.71 & 0.61 & 0.42 & 0.34 \\
 \hline
 \multirow{4}{*}{\shortstack{LOTS-EXIF-Net\\(No Mask)}} & 23 & 0.74 & 0.65 & 0.42 & 0.32 \\
 \cline{2-6} & 30(default) & \bf{0.68} & \bf{0.58} & \bf{0.36} & \bf{0.25} \\
 \cline{2-6} & 37 & 0.75 & 0.67 & 0.44 & 0.35 \\
 \cline{2-6} & 45 & 0.75 & 0.68 & 0.45 & 0.37 \\
 \hline

\end{tabular}
\end{table} 

We conducted an additional experiment to understand how knowledge of the splicing region mask affects LOTS's performance.   To simulate conditions where the attacker does not know the mask, our experiments used the upper-left corner of the image to build target feature maps when no mask was available.  Results of this experiment are shown in Table~\ref{No-Mask}.
From this experiment, we can see that the LOTS-based attack performance drops once not given the mask if the analysis block grid is not aligned with the attack block grid.  For example, on the DSO-1 dataset the F1 score increased from 
0.36 to 0.45 and the MCC score increased from 0.25 to 0.37 when the number of sampling patches increases to 45 and no mask is used.   
These changes correspond to a $25\%$ increase in the F1 score and a $48\%$ increase in the MCC score, comparing with a $14\%$ increase in the F1 score and a $26\%$ increase in the MCC score if the attacker has the mask.

From these results, we can see that if the LOTS attacks does not have access to the splicing mask and the analysis block grid, it's attack performance can drop a noticeable amount.   By contrast, our proposed attack does not need knowledge of the mask or blocking grid to launch a successful attack.  As a result, our attack does not encounter problems that LOTS does requiring side information that may not be available.

\section{Potential Threats}
\label{sec:threats}

In the previous experiments, we showed that our attack can successfully hide  evidence of an image splicing and prevent an investigator from detecting and localizing the fake content.
In this section, we show that our proposed attack can also be used in a different way to enable a new misinformation threat. Specifically, our attack can make real content in an image appear convincingly fake. 
As a result, an analyst cannot distinguish between the authentic images and attacked images..

In practice, the attacker can use the attack to modify the forensic traces of a real object in an image.
The object that is attacked will show a convincing different forensic traces than other regions.  In such case the forensic algorithms will identify the attacked object convincingly to be from another source. At the same time, the attacker can claim that the original was anti-forensically attacked to cover up the splicing evidence. Then it becomes difficult for an analysis to tell the authenticity of the altered and unaltered images. This can be used as a misinformation source to cast doubt on a real image

Alternatively, a naive attack could be launched by adding noise or some other content-preserving local editing operation, to make forensic algorithms predict the attacked content to be not real. However, subsequent analysis would likely show that the content has been post-processed, instead of spliced. As a result, the analysis will easily conclude that the image is a modified version of a real image.

This demonstrates that our attack enables a new type of misinformation threat, in which an attacker can take advantage of forensic analysis algorithms and convincingly cast doubt on real content.
An example of this attack is shown in Fig~\ref{fig:falsify_pristine}. 
The top row shows a real image with its localization results produced by each algorithms. We note that the detection algorithms correctly identify this image to be authentic, but localization algorithms are force to always produce some heatmap. Since the content is real, these heatmaps make little sense.
The bottom row of Fig~\ref{fig:falsify_pristine} shows the target localization masks of our attack, the attacked image, as well as localization results produced by each forensic algorithm. Here we can clearly see that our attack causes the woman at the left to be misidentified as being spliced.

The above examples shows that our attack could be not only used to hide the manipulation traces in the splicing image, but also make real object convincingly fake under the image forensics investigation.

\section{Conclusion}
\label{sec:conclusion}

In this paper, we proposed a new GAN-based anti-forensic attack that is able to fool state-of-the-art forensic detection and localization algorithms that utilize Siamese neural networks.  This attack works by using an adversarially trained generator to create synthetic forensic traces that are self-consistent throughout the image.  We propose a novel anti-forensic generator architecture and two phase training strategy.  
In Phase 1, we pre-train the generator to attack the camera model classifier. In Phase 2, we train the generator to attack  targeted Siamese networks using a new loss function.  After  training, our attack is launched by simply passing images through the generator.
Through a series of experiments,  we demonstrated that our proposed attack can fool  splicing detection and localization algorithms while maintaining a high image quality.  
Additionally, we found that our attack can transfer to fool %
algorithms unseen during training such as Noiseprint.
Experimental results showed that our proposed attack outperforms other existing attacks including adversarial example attacks and LOTS.  Finally, we showed how this new attack can be used to launch a new misinformation threat in which real content is convincingly made to appear as fake.

\balance
\vspace{1.5em}

\ifCLASSOPTIONcaptionsoff
  \newpage
\fi

\bibliographystyle{IEEEtran}
\bibliography{TIFS}

\begin{thebibliography}{10}
\providecommand{\url}[1]{#1}
\csname url@samestyle\endcsname
\providecommand{\newblock}{\relax}
\providecommand{\bibinfo}[2]{#2}
\providecommand{\BIBentrySTDinterwordspacing}{\spaceskip=0pt\relax}
\providecommand{\BIBentryALTinterwordstretchfactor}{4}
\providecommand{\BIBentryALTinterwordspacing}{\spaceskip=\fontdimen2\font plus
\BIBentryALTinterwordstretchfactor\fontdimen3\font minus
  \fontdimen4\font\relax}
\providecommand{\BIBforeignlanguage}[2]{{%
\expandafter\ifx\csname l@#1\endcsname\relax
\typeout{** WARNING: IEEEtran.bst: No hyphenation pattern has been}%
\typeout{** loaded for the language `#1'. Using the pattern for}%
\typeout{** the default language instead.}%
\else
\language=\csname l@#1\endcsname
\fi
#2}}
\providecommand{\BIBdecl}{\relax}
\BIBdecl

\bibitem{stamm2013overview}
M.~C. Stamm, M.~Wu, and K.~J.~R. Liu, ``Information forensics: An overview of
  the first decade,'' \emph{IEEE Access}, vol.~1, pp. 167--200, 2013.

\bibitem{piva2013overview}
A.~Piva, ``An overview on image forensics,'' \emph{International Scholarly
  Research Notices}, vol. 2013, 2013.

\bibitem{verdoliva2020overview}
L.~Verdoliva, ``Media forensics and deepfakes: An overview,'' \emph{IEEE
  Journal of Selected Topics in Signal Processing}, vol.~14, no.~5, pp.
  910--932, 2020.

\bibitem{bayar2018constrained}
B.~Bayar and M.~C. Stamm, ``Constrained convolutional neural networks: A new
  approach towards general purpose image manipulation detection,'' \emph{IEEE
  Transactions on Information Forensics and Security}, vol.~13, no.~11, pp.
  2691--2706, 2018.

\bibitem{chen2015median}
J.~Chen, X.~Kang, Y.~Liu, and Z.~J. Wang, ``Median filtering forensics based on
  convolutional neural networks,'' \emph{IEEE Signal Processing Letters},
  vol.~22, no.~11, pp. 1849--1853, 2015.

\bibitem{cozzolino2017recasting}
D.~Cozzolino, G.~Poggi, and L.~Verdoliva, ``Recasting residual-based local
  descriptors as convolutional neural networks: an application to image forgery
  detection,'' in \emph{ACM Workshop on Information Hiding and Multimedia
  Security}, 2017, pp. 159--164.

\bibitem{BARNI2017153}
M.~Barni, L.~Bondi, N.~Bonettini, P.~Bestagini, A.~Costanzo, M.~Maggini,
  B.~Tondi, and S.~Tubaro, ``Aligned and non-aligned double jpeg detection
  using convolutional neural networks,'' \emph{Journal of Visual Communication
  and Image Representation}, vol.~49, pp. 153--163, 2017.

\bibitem{bondi2016first}
L.~Bondi, L.~Baroffio, D.~G{\"u}era, P.~Bestagini, E.~J. Delp, and S.~Tubaro,
  ``First steps toward camera model identification with convolutional neural
  networks,'' \emph{IEEE Signal Processing Letters}, vol.~24, no.~3, pp.
  259--263, 2016.

\bibitem{bayar2017design}
B.~Bayar and M.~C. Stamm, ``Design principles of convolutional neural networks
  for multimedia forensics,'' \emph{Electronic Imaging}, vol. 2017, no.~7, pp.
  77--86, 2017.

\bibitem{tuama2016camera}
A.~Tuama, F.~Comby, and M.~Chaumont, ``Camera model identification with the use
  of deep convolutional neural networks,'' in \emph{2016 IEEE International
  workshop on information forensics and security (WIFS)}.\hskip 1em plus 0.5em
  minus 0.4em\relax IEEE, 2016, pp. 1--6.

\bibitem{Amerini_2017_WIFS}
I.~Amerini, T.~Uricchio, and R.~Caldelli, ``Tracing images back to their social
  network of origin: A cnn-based approach,'' in \emph{IEEE Workshop on
  Information Forensics and Security (WIFS)}, 2017, pp. 1--6.

\bibitem{cozzolino2019noiseprint}
D.~Cozzolino and L.~Verdoliva, ``Noiseprint: A cnn-based camera model
  fingerprint,'' \emph{IEEE Transactions on Information Forensics and
  Security}, vol.~15, pp. 144--159, 2019.

\bibitem{mayer2020exposing}
O.~Mayer and M.~C. Stamm, ``Exposing fake images with forensic similarity
  graphs,'' \emph{IEEE Journal of Selected Topics in Signal Processing},
  vol.~14, no.~5, pp. 1049--1064, 2020.

\bibitem{huh2018fighting}
M.~Huh, A.~Liu, A.~Owens, and A.~A. Efros, ``Fighting fake news: Image splice
  detection via learned self-consistency,'' in \emph{Proc. of the European
  Conference on Computer Vision (ECCV)}, 2018, pp. 101--117.

\bibitem{stamm2010anti}
M.~C. Stamm, S.~K. Tjoa, W.~S. Lin, and K.~R. Liu, ``Anti-forensics of jpeg
  compression,'' in \emph{2010 IEEE International Conference on Acoustics,
  Speech and Signal Processing}.\hskip 1em plus 0.5em minus 0.4em\relax IEEE,
  2010, pp. 1694--1697.

\bibitem{fan2014jpeg}
W.~Fan, K.~Wang, F.~Cayre, and Z.~Xiong, ``Jpeg anti-forensics with improved
  tradeoff between forensic undetectability and image quality,'' \emph{IEEE
  Transactions on Information Forensics and Security}, vol.~9, no.~8, pp.
  1211--1226, 2014.

\bibitem{Kirchner_2008_TIFS}
M.~Kirchner and R.~Bohme, ``Hiding traces of resampling in digital images,''
  \emph{IEEE Transactions on Information Forensics and Security}, vol.~3,
  no.~4, pp. 582--592, 2008.

\bibitem{wu2013anti}
Z.-H. Wu, M.~C. Stamm, and K.~R. Liu, ``Anti-forensics of median filtering,''
  in \emph{2013 IEEE International Conference on Acoustics, Speech and Signal
  Processing}.\hskip 1em plus 0.5em minus 0.4em\relax IEEE, 2013, pp.
  3043--3047.

\bibitem{fontani2012hiding}
M.~Fontani and M.~Barni, ``Hiding traces of median filtering in digital
  images,'' in \emph{2012 Proceedings of the 20th European Signal Processing
  Conference (EUSIPCO)}.\hskip 1em plus 0.5em minus 0.4em\relax IEEE, 2012, pp.
  1239--1243.

\bibitem{Owen2015antilca}
O.~Mayer and M.~C. Stamm, ``{Anti-forensics of chromatic aberration},'' in
  \emph{Media Watermarking, Security, and Forensics 2015}, A.~M. Alattar, N.~D.
  Memon, and C.~D. Heitzenrater, Eds., vol. 9409, International Society for
  Optics and Photonics.\hskip 1em plus 0.5em minus 0.4em\relax SPIE, 2015, pp.
  192 -- 200.

\bibitem{Akhtar2018survey}
N.~Akhtar and A.~Mian, ``Threat of adversarial attacks on deep learning in
  computer vision: A survey,'' \emph{IEEE Access}, vol.~6, pp.
  14\,410--14\,430, 2018.

\bibitem{goodfellow2015explain}
I.~Goodfellow, J.~Shlens, and C.~Szegedy, ``Explaining and harnessing
  adversarial examples,'' in \emph{International Conference on Learning
  Representations}, 2015.

\bibitem{madry2018towards}
A.~Madry, A.~Makelov, L.~Schmidt, D.~Tsipras, and A.~Vladu, ``Towards deep
  learning models resistant to adversarial attacks,'' in \emph{International
  Conference on Learning Representations}, 2018.

\bibitem{carlini2017towards}
N.~Carlini and D.~Wagner, ``Towards evaluating the robustness of neural
  networks,'' in \emph{2017 IEEE Symposium on Security and Privacy (sp)}.\hskip
  1em plus 0.5em minus 0.4em\relax IEEE, 2017, pp. 39--57.

\bibitem{Barni_2019_ICASSP}
M.~Barni, K.~Kallas, E.~Nowroozi, and B.~Tondi, ``On the transferability of
  adversarial examples against cnn-based image forensics,'' in \emph{ICASSP
  2019 - 2019 IEEE International Conference on Acoustics, Speech and Signal
  Processing (ICASSP)}, 2019, pp. 8286--8290.

\bibitem{Carlini2020CVPRW}
N.~Carlini and H.~Farid, ``Evading deepfake-image detectors with white- and
  black-box attacks,'' in \emph{IEEE/CVF Conference on Computer Vision and
  Pattern Recognition Workshops (CVPRW)}, 2020, pp. 2804--2813.

\bibitem{rozsa2020adversarial}
A.~Rozsa, Z.~Zhong, and T.~E. Boult, ``Adversarial attack on deep
  learning-based splice localization,'' in \emph{Proceedings of the IEEE/CVF
  Conference on Computer Vision and Pattern Recognition Workshops}, 2020, pp.
  648--649.

\bibitem{Guera2017CVPRW}
D.~Güera, Y.~Wang, L.~Bondi, P.~Bestagini, S.~Tubaro, and E.~J. Delp, ``A
  counter-forensic method for cnn-based camera model identification,'' in
  \emph{2017 IEEE Conference on Computer Vision and Pattern Recognition
  Workshops (CVPRW)}, 2017, pp. 1840--1847.

\bibitem{Zhong_2021_TIFS}
Y.~Zhong and W.~Deng, ``Towards transferable adversarial attack against deep
  face recognition,'' \emph{IEEE Transactions on Information Forensics and
  Security}, vol.~16, pp. 1452--1466, 2021.

\bibitem{Chen_2018_ICIP}
C.~Chen, X.~Zhao, and M.~C. Stamm, ``Mislgan: An anti-forensic camera model
  falsification framework using a generative adversarial network,'' in
  \emph{IEEE International Conference on Image Processing (ICIP)}, 2018, pp.
  535--539.

\bibitem{Chen_2019_TIFS}
------, ``Generative adversarial attacks against deep-learning-based camera
  model identification,'' \emph{IEEE Transactions on Information Forensics and
  Security}, pp. 1--1, 2019.

\bibitem{Xie_2021_TCSVT}
H.~Xie, J.~Ni, and Y.-Q. Shi, ``Dual-domain generative adversarial network for
  digital image operation anti-forensics,'' \emph{IEEE Transactions on Circuits
  and Systems for Video Technology}, vol.~32, no.~3, pp. 1701--1706, 2022.

\bibitem{Kim_2018_SPL}
D.~Kim, H.-U. Jang, S.-M. Mun, S.~Choi, and H.-K. Lee, ``Median filtered image
  restoration and anti-forensics using adversarial networks,'' \emph{IEEE
  Signal Processing Letters}, vol.~25, no.~2, pp. 278--282, 2018.

\bibitem{Ding_2021_TM}
F.~Ding, G.~Zhu, Y.~Li, X.~Zhang, P.~K. Atrey, and S.~Lyu, ``Anti-forensics for
  face swapping videos via adversarial training,'' \emph{IEEE Transactions on
  Multimedia}, pp. 1--1, 2021.

\bibitem{Mayer2020TIFS}
O.~Mayer and M.~C. Stamm, ``Forensic similarity for digital images,''
  \emph{IEEE Transactions on Information Forensics and Security}, vol.~15, pp.
  1331--1346, 2020.

\bibitem{mayer2018transfer}
O.~Mayer, B.~Bayar, and M.~C. Stamm, ``Learning unified deep-features for
  multiple forensic tasks,'' in \emph{Proceedings of the 6th ACM Workshop on
  Information Hiding and Multimedia Security}, ser. IH\&MMSec '18.\hskip 1em
  plus 0.5em minus 0.4em\relax Association for Computing Machinery, 2018, p.
  79–84.

\bibitem{Zhang2019wifs}
X.~Zhang, S.~Karaman, and S.-F. Chang, ``Detecting and simulating artifacts in
  gan fake images,'' in \emph{2019 IEEE International Workshop on Information
  Forensics and Security (WIFS)}, 2019, pp. 1--6.

\bibitem{he2016deep}
K.~He, X.~Zhang, S.~Ren, and J.~Sun, ``Deep residual learning for image
  recognition,'' in \emph{Proceedings of the IEEE conference on computer vision
  and pattern recognition}, 2016, pp. 770--778.

\bibitem{hsu06ICME}
Y.-F. Hsu and S.-F. Chang, ``Detecting image splicing using geometry invariants
  and camera characteristics consistency,'' in \emph{International Conference
  on Multimedia and Expo}, 2006.

\bibitem{Carvalho_2013_TIFS}
T.~J. de~Carvalho, C.~Riess, E.~Angelopoulou, H.~Pedrini, and
  A.~de~Rezende~Rocha, ``Exposing digital image forgeries by illumination color
  classification,'' \emph{IEEE Transactions on Information Forensics and
  Security}, vol.~8, no.~7, pp. 1182--1194, 2013.

\bibitem{Korus2016TIFS}
P.~Korus and J.~Huang, ``Multi-scale analysis strategies in prnu-based
  tampering localization,'' \emph{IEEE Trans. on Information Forensics \&
  Security}, 2017.

\bibitem{Rozsa2017LOTS}
A.~Rozsa, M.~Günther, and T.~E. Boult, ``Lots about attacking deep features,''
  in \emph{2017 IEEE International Joint Conference on Biometrics (IJCB)},
  2017, pp. 168--176.

\bibitem{papernot2018cleverhans}
\protect{N. Papernot et al.}, ``Technical report on the cleverhans v2.1.0
  adversarial examples library,'' \emph{arXiv preprint arXiv:1610.00768}, 2018.

\end{thebibliography}

\end{document}